\documentclass[12pt]{article}
\usepackage{amsmath,amssymb}
\usepackage{graphicx}
\allowdisplaybreaks[1]

\makeatletter
 
  \@addtoreset{equation}{section}
 \makeatother

\newcommand{\bpsi}{\bar{\psi}}

\newcommand{\blambda}{\bar{\lambda}}

\newcommand{\hmu}{\hat{\mu}}
\newcommand{\hnu}{\hat{\nu}}

\newcommand{\hPhi}{\hat{\Phi}}

\newcommand{\tA}{{\widetilde{A}}}

\newcommand{\tH}{{\widetilde{H}}}

\newcommand{\tdelta}{\tilde{\delta}}

\newcommand{\cD}{{\cal D}}
\newcommand{\cF}{{\cal F}}

\newcommand{\cM}{{\cal M}}
\newcommand{\cN}{{\cal N}}
\newcommand{\cO}{{\cal O}}

\newcommand{\R}{\mathbb{R}}

\newcommand{\nn}{\nonumber}

\newcommand{\Tr}{{\rm Tr}\,}

\newcommand{\del}{\partial}
\newcommand{\e}{\epsilon}

\newcommand{\be}{\begin{equation}}
\newcommand{\ee}{\end{equation}}
\newcommand{\bea}{\begin{eqnarray}}
\newcommand{\eea}{\end{eqnarray}}

\allowdisplaybreaks
\begin{document}
\begin{titlepage}

\setcounter{page}{0}
\renewcommand{\thefootnote}{\fnsymbol{footnote}}

\begin{flushright}
OIQP-11-07\\
\end{flushright} 
\vspace{10mm}

\begin{center}
{\large\bf
Non-perturbative construction of 2D and 4D
supersymmetric Yang-Mills theories with 8 supercharges 
}

\vspace{10mm}
Masanori Hanada$^{1}$,%
\footnote{\tt mhanada@u.washington.edu}  
So Matsuura$^{2}$,%
\footnote{\tt s.matsu@phys-h.keio.ac.jp}
and Fumihiko Sugino ${}^{3}$
\footnote{\tt fumihiko\_sugino@pref.okayama.lg.jp} 
\vspace{10mm}

{\em 
$^{1}$ Department of Physics, University of Washington, 
Seattle, WA 98195-1560, USA\\
$^{2}$ Department of Physics, 
and Research and Education Center for Natural Science, 
Keio University, 4-1-1 Hiyoshi, Yokohama, 223-8521, Japan\\
$^{3}$ Okayama Institute for Quantum Physics, 
Kyoyama 1-9-1, Kita-ku, Okayama 700-0015, Japan}
\end{center}

\vspace{20mm}
\centerline{{\bf Abstract}}
\vspace{3mm}

In this paper, we consider two-dimensional $\cN=(4,4)$ supersymmetric
Yang-Mills (SYM) theory and deform it by a mass parameter $M$
with keeping all supercharges. 
We further add another mass parameter $m$ in a manner to respect 
two of the eight supercharges and put the deformed theory 
on a two-dimensional square lattice, on which 
the two supercharges are exactly preserved. 
The flat directions of scalar fields are stabilized 
due to the mass deformations, 
which gives discrete minima representing fuzzy spheres. 
We show 
in the perturbation theory 
that the lattice continuum limit can be taken 
without any fine tuning. 
Around the trivial minimum, this lattice theory 
serves as a non-perturbative definition of two-dimensional $\cN=(4,4)$ 
SYM theory. 
We also discuss that the same lattice theory realizes four-dimensional $\cN = 2$ $U(k)$ SYM 
on $\R^2\times ({\rm Fuzzy}\ \R^2)$ 
around the minimum of $k$-coincident fuzzy spheres. 
\end{titlepage}
\newpage

\section{Introduction}

Supersymmetric gauge theories play very important roles 
in theoretical particle physics. 
They are not only promising candidates of 
physics beyond the standard model, 
which are one of the targets in Large Hadron Collider (LHC) experiments, 
but also provide crucial insights into non-perturbative aspects of superstring/M theory 
\cite{Banks:1996vh,Ishibashi:1996xs,Dijkgraaf:1997vv,Maldacena:1997re}. 
Although they are analytically more controllable than non-supersymmetric theories, 
there have arisen many intriguing features conjectured from various duality arguments which cannot be addressed by  
current analytic techniques. Therefore it is important to find promising numerical frameworks 
which enable us to examine them and to obtain new insights into non-perturbative dynamics.   
However, it is not a straightforward task in lattice field theory because of the notorious 
difficulties of lattice supersymmetry. 
So far, 
for one- and two-dimensional 
theories \cite{Catterall:2009it} and ${\cal N}=1$ pure super Yang-Mills (SYM) theories in three \cite{Maru:1997kh} 
and four dimensions \cite{Giedt:2009yd}, 
some lattice models are shown to be free from any fine tuning at least
perturbatively. 
(Lattice models for three- and four-dimensional SYM theories are
constructed by
orbifold or twisting methods preserving some of supersymmetries \cite{Catterall:2009it},
although they
require fine tuning in taking the continuum limit.)
In order to overcome this difficulty, one possible 
direction is to pursue new discretization methods 
different from conventional lattice.

{}For one-dimensional theory (matrix quantum mechanics) a powerful ``non-lattice'' 
technique \cite{Hanada:2007ti}
is applicable.  Maximally supersymmetric matrix quantum mechanics has been studied extensively, and 
remarkable quantitative agreement with the gauge/gravity duality conjecture has been obtained 
\cite{Anagnostopoulos:2007fw,Hanada:2009ne}. 
(Qualitatively consistent results are obtained also from lattice simulation \cite{Catterall:2008yz}.)~\footnote{ 
In particular the simulation results are consistent with the existence of the threshold bound state \cite{Hanada:2009ne}, 
which is an important ingredient of the Matrix theory conjecture \cite{Banks:1996vh}.  
Theories with less supersymmetry have also been studied \cite{Hanada:2010jr} 
and the result strongly suggests the threshold bound state 
does not exist in those cases as expected from the calculation of the Witten index.
For $SU(2)$ theory with four supercharges
the energy spectrum has been studied in \cite{Campostrini:2004bs} by using the Hamiltonian approach. 
The simulation results  \cite{Hanada:2010jr} look consistent with the spectrum calculated in \cite{Campostrini:2004bs}. 
}
{}For two-dimensional $\cN=(2,2)$ SYM, non-perturbative evidences
for the lattice model presented in \cite{Sugino:2004qd} to require 
no fine tuning have been given by numerical simulation for the gauge 
group $G = SU(2)$ in \cite{Kanamori:2008bk} and for
$G = SU(N)$ with $N=2,3,4,5$ in \cite{Hanada:2009hq}. 
{}Furthermore, \cite{Hanada:2010qg} has shown that the model constructed 
in \cite{Cohen:2003qw} is free from the sign
problem and gives the same physics as that in 
\cite{Sugino:2004qd} after an appropriate treatment of the overall
$U(1)$ modes. 
$\cN=(8,8)$ theory has also been simulated in \cite{Catterall:2010fx} 
in order to study the black hole/black string transition. 
(For other numerical studies in the context of the gauge/gravity duality, see e.g. \cite{Berenstein:2008jn,Berenstein:2010xw}.) 
Combining the non-lattice or lattice techniques   
with matrix model/non-commutative space techniques \cite{Berenstein:2002jq,Myers:1999ps}, 
three-dimensional theory can be obtained 
as a theory on fuzzy sphere \cite{Maldacena:2002rb}. 
Also, in the planar limit, four-dimensional theory 
can be obtained using a novel large-$N$ reduction technique \cite{Ishii:2008ib,Ishiki:2011ct} 
inspired by the Eguchi-Kawai equivalence \cite{Eguchi:1982nm}. 
However, four-dimensional theories of extended supersymmetry 
at a finite rank of a gauge group 
were out of reach.\footnote{
Number of fine tunings in the lattice model has been estimated in \cite{Catterall:2011pd}. 
} 

Recently we proposed a new regularization scheme 
for four-dimensional ${\cal N}=4$ SYM with $G=U(k)$, 
which is free from fine tuning \cite{Hanada:2010kt,Hanada:2010gs}. 
It is a hybrid of two-dimensional lattice \cite{Sugino:2003yb,Sugino:2004qd,Sugino:2004uv} 
(see also \cite{Cohen:2003xe,Cohen:2003qw,Kaplan:2005ta,Catterall:2004np,D'Adda:2005zk,Suzuki:2005dx,Sugino:2006uf}) 
and matrix model techniques \cite{Maldacena:2002rb}. 
We regularized 
a two-dimensional SYM with plane wave deformation, which has a fuzzy sphere 
classical solution, and two additional {\it non-commutative} dimensions are 
generated by the fuzzy sphere.\footnote{
Refs.~\cite{Unsal:2005us,Ydri:2007ua} discuss similar construction of four-dimensional non-commutative spaces 
from zero-dimensional matrix models.
}
In four-dimensional ${\cal N}=4$ theory, the commutative limit of 
non-commutative space is believed to be smooth 
\cite{Hashimoto:1999ut, Maldacena:1999mh,Matusis:2000jf}.  
Therefore it is expected that we can numerically simulate 
four-dimensional $\cN=4$ SYM on $\R^4$ using this formulation.  
 
In this paper we provide a non-perturbative formulation of two-dimensional 
$\cN=(4,4)$ SYM and 
four-dimensional ${\cal N}=2$ SYM on non-commutative space, 
which is analogous to the one given in \cite{Hanada:2010kt}.
We first deform the action of two-dimensional $\cN=(4,4)$ SYM
by a mass parameter $M$ with keeping {\em all the supersymmetry}. 
The gauge group is $U(N)$ or $SU(N)$. 
As a result of this deformation, flat directions of three scalar fields are stabilized, and 
not only the trivial configuration 
but also fuzzy sphere configurations become supersymmetric classical 
solutions. 
We further introduce another mass parameter $m$, which keeps  
two supercharges, $Q_\pm$,  
in order to lift up the flat direction of the remaining scalar field. 
Then the deformed theory is formulated on a two-dimensional square lattice 
in a manner to preserve $Q_\pm$ exactly. 
Here we use a prescription developed by one of the authors, 
{}F.S. \cite{Sugino:2003yb,Sugino:2004qd}.
Thanks to the deformations by $M$ and $m$, 
we can solve the problem of 
the running of the vacuum expectation values of the scalar fields along the flat directions. 
In this sense, the formulation in this paper can be regarded as a modification of the formulation 
in \cite{Sugino:2004qd} to stabilize all the flat directions of scalar fields  
with keeping $Q_\pm$ supersymmetries.  
We will give a perturbative argument that the continuum limit of the 
two-dimensional lattice does not require any fine tuning. 
Note that the deformation by $m$ does not affect 
the fuzzy sphere solutions. 
Namely, 
they are still solutions preserving $Q_\pm$ in the lattice theory 
even after introducing $m$.  
We next consider the lattice theory with the gauge group $U(N)$ expanded around a $k$-coincident 
fuzzy sphere background. 
Taking the lattice continuum limit first, 
we obtain four-dimensional $\cN=2$ SYM on $\R^2\times ({\rm Fuzzy}\ S^2)$
with the gauge group $U(k)$ 
deformed by the parameter $m$.
Here the matrix size is given by $N=kn$, 
the radius of the fuzzy spheres $R=\frac{3}{M}$, 
the noncommutativity parameter $\Theta=\frac{18}{M^2n}$, 
the four-dimensional gauge coupling $g_{4d}^2 = 2\pi\Theta g_{2d}^2$, 
and the naturally introduced UV cutoff $\frac{M}{3}(n-1)$.   
Although the preserved supercharges at this stage are only 
$Q_\pm$, the supersymmetry breaking is soft by the mass parameter $m$. 
Therefore all the supersymmetry is recovered by 
simply taking the 
limit of $m\to 0$. 
{}Finally, we take a large-$N$ limit sending $M\to 0$ with fixing 
$k$ and $\Theta$. 
In this limit, the fuzzy sphere becomes non-commutative Moyal 
plane $\R_\Theta^2$. 
Because of the full supersymmetry preserved upon turning off $M$, we can strongly expect that 
four-dimensional $\cN=2$ SYM on $\R^2\times \R_\Theta^2$
is realized  
without any fine tuning. 


This paper is organized as follows. 
In the next section, we review continuum $\cN=(4,4)$ 
SYM theory in two dimensions 
and rewrite the action in the so-called balanced topological field theory (BTFT) form. 
In the section 3, we add appropriate terms 
depending on the parameter $M$ to the action, so that all the supercharges are preserved. 
Furthermore, the mass $m$ is introduced to stabilize all the flat directions. 
In the section 4, we put the deformed theory on a two-dimensional 
lattice in a manner to keep two supercharges. 
In the section 5, we present an intriguing scenario leading to four-dimensional 
$\cN=2$ SYM with a finite-rank gauge group $U(k)$, 
starting with the two-dimensional lattice formulation given in the
section 4. 
The section 5 is devoted to conclusion and discussion. 
In the appendix A, we explain how to derive the mass deformation by $M$ that keeps 
all the supercharges. 
In the appendix B, the explicit form of the lattice action is presented. 

\section{Continuum two-dimensional $\cN=(4,4)$ supersymmetric Yang-Mills
 theory}
In this section we recast 
the two-dimensional $\cN=(4,4)$ SYM theory on (Euclidean) $\R^2$ with gauge group $G=U(N)$ or $SU(N)$ 
to a convenient form for our lattice formulation (BTFT form \cite{Dijkgraaf:1996tz}).   
The action of the theory reads 
\begin{align}
 S_{2d}=\frac{2}{g^2_{2d}}\int d^2x\, \Tr \Bigl[
&\frac{1}{2}F_{12}^2+\frac{1}{2}\left(\cD_\mu X^I\right)^2
-\frac{1}{4}[X^I,X^J]^2 \nn \\
&+\frac{1}{2}\Psi^T (\cD_1+\gamma_2 \cD_2)\Psi 
+\frac{i}{2}\Psi^T \gamma_I[X^I,\Psi]
\Bigr], 
\label{2dSYM}
\end{align}
where $\mu=1,2$, $I=3,\cdots,6$,
$\cD_\mu=\del_\mu+i[A_\mu,\cdot]$, and  
all the fields are in the adjoint representation of $G$.  
They are expanded by a basis of the representation $T^a$ ($a=1, \cdots, \dim (G)$) 
as $A_\mu=A_\mu^a T^a, \cdots$.  
The fermion $\Psi$ is an 8-component spinor which is real, $\Psi^{a*}=\Psi^a$, 
but not Majorana. 
The gamma matrices 
$\gamma_i$ ($i=2,\cdots,6$) are
$8\times 8$ matrices satisfying $\{\gamma_i,\gamma_j\}=-2\delta_{ij}$ 
and $\gamma_2\cdots\gamma_6=-i {\bf 1}_8$.  
Their explicit form we use is 
\begin{align}
 \gamma_2 &= -i\left(\begin{matrix}
\sigma_3&&&\\
&\sigma_3&&\\
&&\sigma_3&\\
&&&\sigma_3\end{matrix}\right), \quad 
\gamma_3=\left(\begin{matrix}
&-\sigma_2&& \\
\sigma_2&&& \\
&&&-\sigma_2 \\
&&\sigma_2&
\end{matrix}\right), \nn \\ 
\gamma_4&=-i\left(\begin{matrix}
\sigma_1&&& \\
&\sigma_1&& \\
&&\sigma_1& \\
&&&\sigma_1
\end{matrix}\right), \quad 
\gamma_5=\left(\begin{matrix}
&&-\sigma_2& \\
&&&\sigma_2 \\\
\sigma_2&&& \\
&-\sigma_2&&
\end{matrix}\right), \quad 
\gamma_6=\left(\begin{matrix}
&&&-\sigma_2 \\
&&-\sigma_2& \\
&\sigma_2&& \\
\sigma_2&&&
\end{matrix}\right), 
\label{gamma}
\end{align}
where $\sigma_{1,2,3}$ are Pauli matrices. 
This theory preserves eight supercharges and 
the supersymmetry transformation of the fields is given by 
\begin{align}
 \delta' A_1&= \epsilon^T \Psi, \quad
 \delta' A_2 = \epsilon^T \gamma_2 \Psi, \quad
 \delta' X^I = \e^T \gamma_I \Psi, \nn \\
 \delta' \Psi &= \left(-F_{12}\gamma_2 -(\cD_1 X^I)\gamma_I
+(\cD_2 X^I)\gamma_{2I}+\frac{i}{2}[X^I,X^J]\gamma_{IJ}\right) \e, 
\label{susy trans}
\end{align} 
where $\e$ is an 8-component supersymmetry transformation parameter.

As a preparation to construct a lattice theory later, 
we transcribe the theory in terms of topologically twisted variables. 
The two-dimensional $\cN=(4,4)$ SYM has three global 
$U(1)$ symmetries: 
the rotation of $(x^1,x^2)$-plane $U(1)_E$, 
the R-symmetry $U(1)_R$, 
and another rotation $U(1)_V$ whose origin is  
the chiral rotation in four-dimensional $\cN=2$ SYM from the viewpoint of dimensional reduction. 
We write the fermion $\Psi$ as 
\begin{equation}
 \Psi=\sqrt{2}\left(\xi_R^1,\xi_L^1,\zeta_R^1,\zeta_L^1,
\xi_R^2, \xi_L^2, \zeta_R^2, \zeta_L^2\right)^T,
\end{equation}
and define the following complex combinations, 
\begin{align}
 \lambda_R&\equiv \xi_R^1+i\zeta_R^1, \quad
 \blambda_R\equiv \xi_R^1-i\zeta_R^1, \quad 
 \lambda_L\equiv \xi_L^1+i\zeta_L^1, \quad
 \blambda_L\equiv \xi_L^1-i\zeta_L^1, \nn \\
 \psi_R&\equiv \xi_R^2+i\zeta_R^2, \quad
 \bpsi_R\equiv \xi_R^2-i\zeta_R^2, \quad
 \psi_L\equiv \xi_L^2+i\zeta_L^2, \quad
 \bpsi_L\equiv \xi_L^2-i\zeta_L^2.  
\end{align}
The $U(1)$-charges of the fields are summarized as 
\begin{center}
\begin{tabular}{|c|c|c|c|c|}
 \hline
 {}Fields & $U(1)_E$  & $U(1)_V$ & $U(1)_d$ & $U(1)_R$ \\ \hline \hline
 $A_1\mp iA_2$ & $\pm 1$ & $0$ & $\pm1$ & $0$ \\ 
 $X^3,\ X^4$ & $0$ & $0$ & $0$ & $0$ \\
 $X^5\pm iX^6$ & $0$ & $0$ &$0$ & $\pm 2$ \\
 $\lambda_R$ & $-\frac{1}{2}$ & $\frac{1}{2}$ &$0$ & $1$ \\
 $\lambda_L$ & $\frac{1}{2}$ & $\frac{1}{2}$ &$1$ & $1$ \\
 $\blambda_R$ & $-\frac{1}{2}$ & $-\frac{1}{2}$ &$-1$ & $-1$ \\
 $\blambda_L$ & $\frac{1}{2}$ & $-\frac{1}{2}$ &$0$ & $-1$ \\
 $\psi_R$ & $-\frac{1}{2}$ & $-\frac{1}{2}$ &$-1$ & $1$ \\
 $\psi_L$ & $\frac{1}{2}$ & $-\frac{1}{2}$ &$0$ & $1$ \\
 $\bpsi_R$ & $-\frac{1}{2}$ & $\frac{1}{2}$ &$0$ & $-1$ \\
 $\bpsi_L$ & $\frac{1}{2}$ & $\frac{1}{2}$ &$1$ & $-1$ \\ \hline 
\end{tabular}
\end{center}
where $U(1)_d$ is the diagonal subgroup of $U(1)_E\times U(1)_V$. 
We rename the fields based on the symmetry $U(1)_d$ as 
\begin{align}
 B&\equiv X^3, \quad C\equiv 2X^4, \quad 
 \phi_\pm \equiv X^5\pm iX^6, \nn \\
 \lambda_R &\equiv \frac{1}{\sqrt{2}}
 \left( -\chi_+ + \frac{i}{2}\eta_+ \right), \quad
  \lambda_L \equiv \frac{1}{\sqrt{2}}
 \left( \psi_{+1} -i \psi_{+2} \right), \nn \\
 \blambda_R &\equiv  \frac{1}{\sqrt{2}}
 \left( \psi_{-1}+i\psi_{-2}\right), \quad 
 \blambda_L \equiv  \frac{1}{\sqrt{2}}
 \left( -\chi_-+\frac{i}{2}\eta_- \right), \nn \\
 \psi_R &\equiv  \frac{1}{\sqrt{2}}
 \left( \psi_{+1} +i \psi_{+2} \right), \quad 
 \psi_L \equiv  \frac{1}{\sqrt{2}}
 \left( \chi_+ + \frac{i}{2}\eta_+ \right), \nn \\
 \bpsi_R &\equiv  \frac{1}{\sqrt{2}}
 \left( \chi_- + \frac{i}{2}\eta_- \right), \quad
 \bpsi_L \equiv  \frac{1}{\sqrt{2}}
 \left( \psi_{-1}-i \psi_{-2} \right). 
\label{field redef}
\end{align}
Correspondingly, we define a new expression of the $8$-component 
spinor, 
\begin{align} 
\Psi^{(0)}=\left(\psi_{+1},\psi_{+2},\chi_+,\frac{1}{2}\eta_+, 
\psi_{-1},\psi_{-2},\chi_-,\frac{1}{2}\eta_-\right)^T, 
\end{align}
which is related with $\Psi$ by a unitary transformation, 
\begin{align}
 \Psi&=U_8 \Psi^{(0)}, 
\end{align}
with  
\begin{align}
 U_8&=\frac{1}{2}\left(\begin{matrix}
0&0&-1&i&1&i&0&0\\
1&-i&0&0&0&0&-1&i \\
0&0&i&1&i&-1&0&0 \\
-i&-1&0&0&0&0&-i&-1 \\
1&i&0&0&0&0&1&i \\
0&0&1&i&1&-i&0&0 \\
-i&1&0&0&0&0&i&-1 \\
0&0&-i&1&i&1&0&0
\end{matrix}\right). 
\end{align}

Next we consider the two supercharges, $Q'_+$ and $Q'_-$, 
corresponding to the supersymmetry parameters $\e_+$ and $\e_-$,
\begin{equation}
 \e_\pm^T = \e_\pm^{\prime T} U_8^{-1}, 
 \label{epsilon12}
\end{equation}
with
\begin{equation}
 \e'_+ = \left(\begin{matrix}
\varepsilon_+ \\ 0 \\ 0 \\ 0 \\ 0 \\ 0 \\ 0 \\ 0
\end{matrix}\right), \qquad
 \e'_- = \left(\begin{matrix}
0 \\ 0 \\ 0 \\ 0 \\ \varepsilon_- \\ 0 \\ 0 \\ 0
\end{matrix}\right), 
\quad (\varepsilon_\pm\text{ : Grassmann numbers})
\label{epsilon12-2} 
\end{equation}
respectively. 
The transformation of the fields by $Q'_\pm$ is given by 
\begin{align}
 Q'_\pm A_\mu&=\psi_{\pm\mu}, \quad
 Q'_\pm \psi_{\pm\mu}=\pm i\cD_\mu \phi_\pm, \quad
 Q'_\mp \psi_{\pm\mu}=\frac{i}{2}\cD_\mu C \mp \tH_\mu, \nn \\
 Q'_\pm \tH_\mu &=
 [\phi_\pm,\psi_{\mp\mu}]\mp \frac{1}{2}[C,\psi_{\pm\mu}]
 \mp\frac{i}{2}\cD_\mu \eta_\pm,  \nn \\
 Q'_\pm B &= \chi_\pm, \quad 
 Q'_\pm \chi_\pm = \pm[\phi_\pm,B], \quad
 Q'_\mp \chi_\pm = \frac{1}{2}[C,B] \mp H, \nn \\
 Q'_\pm H &= [\phi_\pm,\chi_\mp] \pm\frac{1}{2}[B,\eta_\pm]
\mp \frac{1}{2}[C,\chi_\pm],
 \nn \\
 Q'_\pm C &= \eta_\pm, \quad
 Q'_\pm \eta_\pm = \pm[\phi_\pm,C], \quad
 Q'_\mp \eta_\pm = \mp[\phi_+,\phi_-], \nn \\
 Q'_\pm \phi_\pm &= 0, \quad
 Q'_\mp \phi_\pm = \mp \eta_\pm,
\end{align}
where $H$ and $\tH_\mu$ are auxiliary fields. 
It is seen that $Q'_\pm$ satisfy 
\begin{align}
 Q_+^{\prime 2}&= \left(\text{infinitesimal gauge transformation with parameter
 $\phi_+$}\right), \nn \\
 Q_-^{\prime 2}&= \left(\text{infinitesimal gauge transformation with parameter
 $-\phi_-$}\right), \nn \\
 \left\{Q'_+,Q'_-\right\} &= 
\left(\text{infinitesimal gauge transformation with parameter
 $C$}\right). 
\end{align}
Then we can rewrite the action (\ref{2dSYM}) in the so-called 
BTFT form \cite{Dijkgraaf:1996tz}; 
\begin{align}
 S_{2d}&= Q'_+ Q'_- \cF, 
\end{align}
with
\begin{align}
 \cF &\equiv \frac{1}{g^2_{2d}}\int d^2x \Tr \biggl[
-iB\Phi -\psi_{+\mu}\psi_{-\mu}-\chi_+\chi_- - \frac{1}{4}\eta_+\eta_-
\biggr], 
\label{BTFT form}
\end{align}
where $\Phi=2 F_{12}$. 
After integrating out the auxiliary fields, the explicit form of the action reads 
\begin{align}
 S_{2d}=&\frac{1}{g_{2d}^2}\int d^2x\, \Tr \biggl[
{}F_{12}^2+(\cD_\mu B)^2+\frac{1}{4}(\cD_\mu C)^2 + \cD_\mu {\phi_+} \cD_\mu
 {\phi_-} \nn \\
&+\frac{1}{4}[{\phi_+},{\phi_-}]^2+[B,{\phi_+}][{\phi_-},B]-\frac{1}{4}[B,C]^2
+\frac{1}{4}[C,{\phi_+}][{\phi_-},C] \nn \\
&+2i\chi_-\left(\cD_1\psi_{+2}-\cD_2\psi_{+1}\right)
-2i\chi_+\left(\cD_1\psi_{-2}-\cD_2\psi_{-1}\right) \nn \\
&+2B\left(\left\{\psi_{+1},\psi_{-2}\right\}-\left\{\psi_{+2},\psi_{-1}\right\}\right)
 \nn \\
&+i\eta_+ \cD_\mu \psi_{-\mu} + i\eta_- \cD_\mu \psi_{+\mu}
-C\left\{\psi_{+\mu},\psi_{-\mu}\right\} \nn \\
&-2\psi_{-\mu}\psi_{-\mu}{\phi_+} -2 \psi_{+\mu}\psi_{+\mu} {\phi_-} \nn \\
&-\chi_-[{\phi_+},\chi_-]+\chi_+[{\phi_-},\chi_+]
+\chi_-[C,\chi_+] -\chi_+[B,\eta_-] -\chi_-[B,\eta_+] \nn \\
&+\frac{1}{4}\eta_+ [{\phi_-},\eta_+]-\frac{1}{4}\eta_-[{\phi_+},\eta_-]
-\frac{1}{4}\eta_+[C,\eta_-]
\biggr]. 
\label{2dSYM-2}
\end{align}

This action is manifestly symmetric under an $SU(2)_R$ subgroup
of the R-symmetry group $SU(4)$. 
The generators of the $SU(2)_R$ are represented as 
\begin{align}
 J_{++}=&\int d^2 x \biggl[
\psi^a_{+\mu}(x)\frac{\delta}{\delta \psi^a_{-\mu}(x)}
+\chi^a_+(x)\frac{\delta}{\delta \chi^a_{-}(x)}
-\eta^a_+(x)\frac{\delta}{\delta\eta^a_-(x)} \nn \\
&+2{\phi_+}^a(x)\frac{\delta}{\delta C^a(x)}
-C^a(x)\frac{\delta}{\delta {\phi_-}^a(x)}
\biggr], \nn \\
 J_{--}=&\int d^2 x \biggl[
\psi^a_{-\mu}(x)\frac{\delta}{\delta \psi^a_{+\mu}(x)}
+\chi^a_-(x)\frac{\delta}{\delta \chi^a_{+}(x)}
-\eta^a_-(x)\frac{\delta}{\delta\eta^a_+(x)} \nn \\
&-2{\phi_-}^a(x)\frac{\delta}{\delta C^a(x)}
+C^a(x)\frac{\delta}{\delta {\phi_+}^a(x)}
\biggr], \nn \\
 J_{0}=&\int d^2 x \biggl[
\psi^a_{+\mu}(x)\frac{\delta}{\delta\psi^a_{+\mu}(x)}
-\psi^a_{-\mu}(x)\frac{\delta}{\delta\psi^a_{^\mu}(x)} 
+\chi^a_+(x)\frac{\delta}{\delta \chi^a_+(x)}
-\chi^a_-(x)\frac{\delta}{\delta \chi^a_-(x)} \nn \\
&+\eta^a_+(x)\frac{\delta}{\delta \eta^a_+(x)}
-\eta^a_-(x)\frac{\delta}{\delta \eta^a_-(x)} 
+2{\phi_+}^a(x)\frac{\delta}{\delta{\phi_+}^a(x)}
-2{\phi_-}^a(x)\frac{\delta}{\delta{\phi_-}^a(x)}
\biggr], 
\end{align}
which satisfy the $SU(2)$ algebra, 
\begin{equation}
 [J_0, J_{\pm\pm}]=\pm2 J_{\pm\pm}, \quad
 [J_{++},J_{--}] = J_0. 
\end{equation}
We see that $(\psi_{+\mu},\psi_{-\mu})$, $(\chi_+,\chi_-)$, 
$(\eta_+, -\eta_-)$ and $(Q'_+, Q'_-)$ transform as doublets 
and $(\phi_+, C, -\phi_-)$ as a triplet under the $SU(2)_R$ 
transformation. 
\section{Mass deformation with keeping 8 supercharges}
We next deform the action (\ref{2dSYM}) by introducing a mass 
parameter $M$;
\begin{align}
 S_{2d,M}=& S_{2d} + S_M, 
\label{2dSYM-M}
\end{align}
with 
\begin{align}
 S_M= &\frac{2}{g_{2d}^2}\int d^2 x \Tr \biggl[
\frac{1}{2}\left(\frac{M}{3}\right)^2(X^p)^2
-i\frac{M}{6}\Psi^T \gamma_{23} \Psi 
+i\frac{M}{3}X^3F_{12}
+i\frac{M}{3}\e_{pqr}X^pX^qX^r
\biggr] \nn \\
= &\frac{1}{g_{2d}^2}\int d^2 x \Tr \biggl[
\frac{M^2}{9}\left(\frac{1}{4}C(x)^2+\phi_+(x)\phi_-(x)\right)
-\frac{M}{2}C(x)\left[\phi_+(x),\phi_-(x)\right] \nn \\
&+i\frac{2M}{3}B(x)F_{12}(x)
+\frac{2M}{3}\psi_{+\mu}(x)\psi_{-\mu}(x)
+\frac{2M}{3}\chi_+(x)\chi_-(x)
-\frac{M}{6}\eta_+(x)\eta_-(x)
\biggr], 
\end{align}
where $p,q,r=4,5,6$. 
This deformation is derived from an eight-supersymmetry 
analogue \cite{Kim:2006wg} 
of the plane wave
matrix model \cite{Berenstein:2002jq}. 
The derivation is summarized in the appendix \ref{app.A}. 
As discussed there, (\ref{2dSYM-M}) is invariant under 
the supersymmetry transformation, 
\begin{equation}
 \delta = \delta'+\delta_M 
 \label{SUSY-M}
\end{equation}
with $\delta'$ given by (\ref{susy trans}) and 
\begin{align}
 \delta_M A_\mu &= \delta_M X^I = 0, \quad 
 \delta_M \Psi = -\frac{M}{3}X^p\gamma_p\gamma_{456}\e. 
\end{align}
Namely, the deformed theory (\ref{2dSYM-M}) still preserves 
eight supercharges. 

In order to rewrite (\ref{2dSYM-M}) in the BTFT form, 
we define the deformed supercharges $Q_\pm$ through the 
deformed supersymmetry transformation (\ref{SUSY-M}) and  
the supersymmetry parameters (\ref{epsilon12}). 
The $Q_\pm$ transformation of the 
fields is~\footnote{
The transformation of the auxiliary fields is 
determined so that relations (\ref{nil-2}) hold. 
} 
\begin{align}
 Q_\pm A_\mu&=\psi_{\pm\mu}, \quad
 Q_\pm \psi_{\pm\mu}=\pm i\cD_\mu \phi_\pm, \quad
 Q_\mp \psi_{\pm\mu}=\frac{i}{2}\cD_\mu C \mp \tH_\mu, \nn \\
 Q_\pm \tH_\mu &=
 [\phi_\pm,\psi_{\mp\mu}]\mp \frac{1}{2}[C,\psi_{\pm\mu}]
 \mp\frac{i}{2}\cD_\mu \eta_\pm +\frac{M}{3}\psi_{\pm\mu},  \nn \\
 Q_\pm B &= \chi_\pm, \quad 
 Q_\pm \chi_\pm = \pm[\phi_\pm,B], \quad
 Q_\mp \chi_\pm = \frac{1}{2}[C,B] \mp H, \nn \\
 Q_\pm H &= [\phi_\pm,\chi_\mp] \pm\frac{1}{2}[B,\eta_\pm]
\mp \frac{1}{2}[C,\chi_\pm] + \frac{M}{3}\chi_\pm,
 \nn \\
 Q_\pm C &= \eta_\pm, \quad
 Q_\pm \eta_\pm = \pm[\phi_\pm,C]+\frac{2M}{3}\phi_\pm, \quad
 Q_\mp \eta_\pm = \mp[\phi_+,\phi_-]\pm\frac{M}{3}C, \nn \\
 Q_\pm \phi_\pm &= 0, \quad
 Q_\mp \phi_\pm = \mp \eta_\pm.
 \label{Qpm-M}
\end{align}
We can check that $Q_\pm$ satisfy the nilpotency relations,%
\begin{align}
 Q_+^{2}&= \left(\text{infinitesimal gauge transformation with parameter
 $\phi_+$}\right)+\frac{M}{3} J_{++}, \nn \\
 Q_-^{2}&= \left(\text{infinitesimal gauge transformation with parameter
 $-\phi_-$}\right)-\frac{M}{3} J_{--}, \nn \\
 \left\{Q_+,Q_-\right\} &= 
\left(\text{infinitesimal gauge transformation with parameter
 $C$}\right)-\frac{M}{3} J_0. 
\label{nil-2}
\end{align}

Using $Q_\pm$, the action (\ref{2dSYM-M}) is expressed as~\footnote{
This kind of deformation is discussed for various SYM models in \cite{Kato:2011yh}.
} 
\begin{align}
 S_{2d,M} = \left(Q_+Q_- - \frac{M}{3}\right)\cF, 
\label{BTFT form 2}
\end{align}
where $\cF$ is identical with (\ref{BTFT form}).
Although $S_{2d,M}$ is not precisely $Q_+Q_-$-exact, 
it is $Q_\pm$-invariant. 
In fact, since $\cF$ is gauge and $SU(2)_R$ invariant, 
\begin{equation}
 J_{\pm\pm}\cF = J_0 \cF = 0,  
\end{equation}
and $(Q_+,Q_-)$ is a doublet of $SU(2)_R$, 
\begin{equation}
 J_{\pm\pm}Q_\mp = Q_\pm, \quad J_0 Q_\pm = \pm Q_\pm, 
\end{equation}
we see 
\begin{align}
 Q_+ S_{2d,M} &= Q_+^2Q_- \cF -\frac{M}{3} Q_+ \cF \nn \\
  &=\frac{M}{3}J_{++}Q_- \cF -\frac{M}{3}Q_+\cF = 0, \nn \\
 Q_- S_{2d,M} &= \left(\{Q_+,Q_-\}Q_- - Q_+Q_-^2\right)\cF 
 -\frac{M}{3} Q_- \cF \nn \\
&= -\frac{M}{3}J_0 Q_- \cF + \frac{M}{3} Q_+ J_{--} \cF
 -\frac{M}{3} Q_- \cF = 0. 
\end{align}

In the next section, we consider a two-dimensional lattice theory 
corresponding to this theory. 
If this theory is naively put on a lattice, 
however, we soon find that it is hard to perform 
a numerical simulation because of the flat direction 
along $B(x)$, which causes running of the scalar field. 
In order to avoid it, we further deform the theory 
by introducing an additional mass term to $\cF$; 
\begin{equation}
 \cF \to \cF + \Delta \cF, 
\end{equation}
with
\begin{equation}
 \Delta\cF  = \frac{1}{g_{2d}^2}\int d^2x \Tr \biggl(
\frac{m}{2} B(x)^2
\biggr), 
\label{delta F}
\end{equation}
where $m$ is a real constant. 
This deformation clearly preserves the supercharges $Q_\pm$. 
After these deformations by $M$ and $m$, the action becomes 
\begin{align}
\label{2dSYM-Mm-Q}
 S_{2d}^{M,m}=
&\left(Q_+Q_- - \frac{M}{3}\right)\left(\cF+\Delta\cF\right) \\
=&\frac{1}{g_{2d}^2}\int d^2x\, \Tr \biggl[
{}F_{12}^2+(\cD_\mu B)^2+\frac{1}{4}(\cD_\mu C)^2 + \cD_\mu {\phi_+} \cD_\mu
 {\phi_-} \nn \\
&+\frac{1}{4}[{\phi_+},{\phi_-}]^2+[B,{\phi_+}][{\phi_-},B]-\frac{1}{4}[B,C]^2
+\frac{1}{4}[C,{\phi_+}][{\phi_-},C] \nn \\
&+\frac{M^2}{9}\left(\frac{1}{4}C(x)^2+\phi_+(x)\phi_-(x)\right)
-\frac{M}{2}C(x)\left[\phi_+(x),\phi_-(x)\right] \nn \\
&-\frac{m}{2}\left(\frac{M}{3}+\frac{m}{2}\right)B(x)^2
+2i\left(\frac{M}{3}+\frac{m}{2}\right)B(x)F_{12}(x) \nn \\
&+2i\chi_-\left(\cD_1\psi_{+2}-\cD_2\psi_{+1}\right)
-2i\chi_+\left(\cD_1\psi_{-2}-\cD_2\psi_{-1}\right) \nn \\
&+2B\left(\left\{\psi_{+1},\psi_{-2}\right\}-\left\{\psi_{+2},\psi_{-1}\right\}\right)
 \nn \\
&+i\eta_+ \cD_\mu \psi_{-\mu} + i\eta_- \cD_\mu \psi_{+\mu}
-C\left\{\psi_{+\mu},\psi_{-\mu}\right\} \nn \\
&-2\psi_{-\mu}\psi_{-\mu}{\phi_+} -2 \psi_{+\mu}\psi_{+\mu} {\phi_-} \nn \\
&-\chi_-[{\phi_+},\chi_-]+\chi_+[{\phi_-},\chi_+]
+\chi_-[C,\chi_+] -\chi_+[B,\eta_-] -\chi_-[B,\eta_+] \nn \\
&+\frac{1}{4}\eta_+ [{\phi_-},\eta_+]-\frac{1}{4}\eta_-[{\phi_+},\eta_-]
-\frac{1}{4}\eta_+[C,\eta_-] \nn \\
&+\frac{2M}{3}\psi_{+\mu}(x)\psi_{-\mu}(x)
+\frac{2M}{3}\chi_+(x)\chi_-(x)
-\frac{M}{6}\eta_+(x)\eta_-(x) 
\biggr]. 
\label{2dSYM-Mm}
\end{align}
We see that $m$ must satisfy 
\begin{equation}
 -\frac{2M}{3} < m < 0, 
 \label{positive cond}
\end{equation}
in order for $B(x)$ to have a positive mass squared.

Looking at the bosonic part of this action, 
we see that there are two types of classical solutions; 
the trivial solution, 
\begin{equation}
C(x)=\phi_\pm(x)=B(x)=0, 
\label{trivial solution}
\end{equation}
and the fuzzy sphere solution, 
\begin{equation}
C(x)=\frac{2M}{3}L_3, \quad 
\phi_\pm(x)=\frac{M}{3}(L_1\pm i L_2), \quad 
B(x) = 0, 
\label{FS solution}
\end{equation} 
where $L_a$ $(a=1,2,3)$ belong to an $N$-dimensional 
(not necessary irreducible) representation 
of $SU(2)$ generators satisfying $[L_a,L_b]=i\e_{abc}L_c$.
Around these solutions,  
there is no flat direction 
because of the mass terms. 
Also, the $Q_\pm$ transformation of $\eta_\pm$, 
\begin{align}
 Q_\pm \eta_\pm = \pm[\phi_\pm,C]+\frac{2M}{3}\phi_\pm, \quad
 Q_\mp \eta_\pm = \mp[\phi_+,\phi_-]\pm\frac{M}{3}C, 
\end{align}
shows that these solutions preserve the $Q_\pm$ supersymmetry. 
We here emphasize that 
the shift of $m$ does not affect 
the fuzzy sphere solution. 
{}Furthermore, 
this solution preserves all the eight supercharges 
of the continuous theory in the limit of $m\to0$,  
as seen from the supersymmetry transformation 
(\ref{dyn_SUSY_2dE}) at $m=0$. 

\section{Lattice formulation for two-dimensional 
$\cN=(4,4)$ supersymmetric Yang-Mills theory}
In this section we put the deformed theory on a two-dimensional
square lattice with lattice spacing $a$. 
In this formulation, the supercharges $Q_\pm$ 
are preserved on the lattice, 
the gauge field is expressed 
as a link variable $U_\mu(x)=e^{iaA_\mu(x)} \in G$ 
as usual lattice gauge theory, 
and all the other lattice fields are defined on sites 
and are made dimensionless by multiplying suitable powers of 
$a$ to the continuum counterparts:
\begin{align}
 &({\rm scalars})^{\rm lat} = a ({\rm scalars})^{\rm cont}, \quad 
 ({\rm fermions})^{\rm lat} = a^{3/2} ({\rm fermions})^{\rm cont}, \nn \\ 
 &({\rm auxiliary \ fields})^{\rm lat} = a^2 ({\rm auxiliary \ fields})^{\rm cont}, \quad 
Q_{\pm}^{\rm lat} = a^{1/2} Q_{\pm}^{\rm cont}.
\end{align}
Also, dimensionless coupling constants on the lattice are 
\begin{equation}
 g_0= a g_{2d}, \quad M_0 = a M, \quad m_0 = am. 
\end{equation}
The supersymmetry transformation is realized as 
\begin{align}
 Q_\pm U_\mu(x) =& i\psi_{\pm\mu}(x)U_\mu(x), \nn \\
 Q_\pm \psi_{\pm\mu}(x)=&i\psi_{\pm\mu}(x) \psi_{\pm\mu}(x) 
 \pm i \cD_\mu \phi_{\pm}(x), 
 \nn \\
 Q_\pm \psi_{\mp\mu}(x)=&\frac{i}{2}\{\psi_{+\mu}(x),\psi_{-\mu}(x)\}
 +\frac{i}{2}\cD_\mu C(x) \pm \tH_\mu(x), \nn \\
 Q_\pm \tH_\mu(x)=&-\frac{1}{2}
 \left[\psi_{\mp\mu}(x),
 \phi_\pm(x)+U_\mu(x) \phi_{\pm}(x+\hmu)U_\mu(x)^\dagger\right] \nn \\
 &\pm\frac{1}{4}\left[\psi_{\pm\mu}(x),
 C(x)+U_\mu(x)C(x+\hmu)U_\mu(x)^\dagger\right] \nn \\
 &\mp\frac{i}{2}\cD_\mu \eta_{\pm}(x) 
 \pm \frac{1}{4}[\psi_{\pm\mu}(x)\psi_{\pm\mu}(x), \psi_{\mp\mu}(x)]
 \nn \\
 &+\frac{i}{2}\left[\psi_{\pm\mu}(x),\tH_\mu(x)\right]+\frac{M_0}{3}\psi_{\pm\mu}(x),
\end{align}
for the lattice fields $U_\mu(x)$, $\psi_{\pm\mu}(x)$ and $\tH_\mu(x)$, 
and transformation of the other fields is the same as the one 
in the continuum theory (\ref{Qpm-M}) 
with the obvious replacement $M\to M_0$. 
Here we have used $\cD_\mu$ as a covariant forward difference 
operator, 
\begin{equation}
 \cD_\mu A(x) \equiv U_\mu(x)A(x+\hmu)U_\mu(x)^\dagger
 -A(x),
\end{equation}
for any adjoint field $A(x)$. 
In order to construct a corresponding lattice action, 
we take the lattice counterpart of $\Phi$ as 
\begin{equation}
 \hPhi(x)=\hat\Phi_{U(N)}(x) \equiv \frac{-i \left( U_{12}(x)-U_{21}(x) \right)}
 {1-\frac{1}{\e^2}|| 1 - U_{12}(x) ||^2}, 
\end{equation}
for $G=U(N)$ and 
\begin{equation}
 \hPhi(x)=\hPhi_{SU(N)}(x) \equiv \hPhi_{U(N)}(x) - \frac{1}{N}\Tr\left(\hPhi_{U(N)}(x)\right){\bf
  1}_N, 
\end{equation}
for $G=SU(N)$.  
Here
$U_{\mu\nu}(x)=U_\mu(x)U_\nu(x+\hmu)U_\mu(x+\hnu)^\dagger U_\nu(x)^\dagger$
is a plaquette variable, $\e$ is a constant satisfying
$0<\e<2$ for $G=U(N)$, and   
$0<\e<2\sqrt{2}$ for $N=2,3,4$ and $0<\e<2\sqrt{N}\sin(\pi/N)$ for
$N\ge5$ for $G=SU(N)$, 
and the norm of a matrix is defined by 
$||A||=\sqrt{\Tr(AA^\dagger)}$ \cite{Sugino:2004qd}.

We then put the two-dimensional theory 
(\ref{2dSYM-Mm-Q}) on a lattice 
using the same form of $\cF$ and $\Delta\cF$ in 
(\ref{BTFT form}) and (\ref{delta F}) together with the trivial replacement 
$\frac{1}{g_{2d}^2}\int d^2 x \to \frac{1}{g_0^2}\sum_x$, 
$M\to M_0$ and $m\to m_0$. 
The obtained $Q_\pm$-invariant lattice action is  
\begin{equation}
 S_{\rm lat} = 
\begin{cases}
\left(Q_+Q_- - \frac{M_0}{3}\right) \left(
\cF_{\rm lat} + \Delta\cF_{\rm lat} \right), & 
|| 1 - U_{12}(x) ||<\e\ {\rm for}\ ^\forall x\\
\infty,  & {\rm otherwise}
\end{cases}
\label{lattice action}
\end{equation}
with
\begin{align}
  \cF_{\rm lat} &\equiv \frac{1}{g^2_0} \sum_x \Tr \biggl[
-iB(x)\hPhi(x) -\psi_{+\mu}(x)\psi_{-\mu}(x)-\chi_+(x)\chi_-(x) 
- \frac{1}{4}\eta_+(x)\eta_-(x)
\biggr], \nn \\
\Delta \cF_{\rm lat} &\equiv \frac{1}{g_0^2}\sum_x \Tr \biggl(
\frac{m_0}{2} B(x)^2
\biggr). 
\end{align}
The explicit expression of the lattice action is 
given in the appendix \ref{app.B}.

\subsection{Absence of flat direction and realization of the physical vacuum}
Let us check that the lattice action 
has the minimum only at the pure gauge configuration 
$U_{12}(x)={\bf 1}_N$ which guarantees
that the weak field expansion 
$U_\mu(x) = 1 + iaA_\mu(x) + \frac{(ia)^2}{2!} A_\mu(x)^2 + \cdots$ 
is allowed in the continuum limit so that the lattice theory 
converges to the desired continuum
theory at the classical level.
After integrating out the auxiliary fields, 
bosonic part of the action $S_{\rm lat}$ takes the form, 
\begin{equation}
 S_{\rm lat}^{(B)} = 
\frac{1}{g_0^2}\sum_x \Tr \biggl[
-\frac{m_0}{2}\left(\frac{M_0}{3}+\frac{m_0}{2}\right) B(x)^2
+i\left(\frac{M_0}{3}+\frac{m_0}{2}\right)B(x)\hPhi(x)
\biggr]
+ S_{\rm PDT}, 
 \label{boson part}
\end{equation}
where $S_{\rm PDT}$ denotes positive (semi-)definite terms 
given by (\ref{PDT}).  
We will treat the second term, which is purely
imaginary, as an operator in the reweighting method, 
and consider the minimum of
the remaining part of $S_{\rm lat}^{(B)}$.
If the condition $-\frac{2M_0}{3}< m_0 <0$, the lattice counterpart of (\ref{positive cond}), is satisfied, 
the mass terms in (\ref{boson part}) fix the minimum at
\begin{equation}
 B(x)=0, 
\end{equation}
which is independent of $S_{\rm PDT}$. 
At this minimum, $S_{\rm PDT}$ becomes 
\begin{align}
S_{\rm PDT} =& \frac{1}{g_0^2}\sum_x \Tr\biggl[
\sum_\mu (\cD_\mu X^p(x))^2 
+\left( i[X^p(x), X^q(x)]+\frac{M_0}{3}\e_{pqr} X^r(x) \right)^2 
\biggr] \nn \\
&+\frac{1}{4g_0^2} \sum_x \frac{\Tr[-(U_{12}(x)-U_{21}(x))^2]}
{\left( 1-\frac{1}{\e^2}|| 1 - U_{12}(x) ||^2 \right)^2}
\label{PDT2}
\end{align}
with (\ref{field redef}) for $p,q,r=4,5,6$. 

Looking at the first line, we see that the trivial solution 
(\ref{trivial solution}) and the fuzzy sphere solution 
(\ref{FS solution}) are still classical solutions 
of the lattice theory by taking into account the replacement $M \to M_0$. 
In the same manner as in the continuum theory, there is no flat direction 
around the solutions; we can perform a stable numerical 
simulation with keeping two supercharges. 
Note that the fuzzy sphere solution plays a crucial role 
to discretize four-dimensional $\cN=2$ SYM in the next section. 

As discussed in \cite{Sugino:2004qd}, in the last term of (\ref{PDT2}) representing
the gauge kinetic term, 
the admissibility condition singles out the trivial minimum 
$U_{12}(x) = {\bf 1}_N$. 
It shows that the lattice action has a stable physical vacuum 
and unphysical degeneracies of vacua do not appear. 

\subsection{Absence of the species doubler}
Let us confirm that there is no species doubler 
in the kinetic terms of this lattice action.
Setting $U_\mu(x)=1$, the kinetic terms for bosons and fermions become 
\begin{align}
\label{boson kinetic}
 S_2^{(B)}=&\frac{1}{g_0^2}\sum_x \Tr \biggl\{
\left(\Delta_{\mu}\phi_+(x)\right)\left(\Delta_\mu\phi_-(x)\right)
+\frac{1}{4}\left(\Delta_\mu C(x)\right)^2 
+\left(\Delta_\mu B(x)\right)^2 \nn \\
&\hspace{2cm}+\frac{M_0^2}{9}\left(\phi_+(x)\phi_-(x)+\frac{1}{4}C(x)^2\right)
-\frac{m_0}{2}\left(\frac{M_0}{2}+\frac{m_0}{2}\right)B(x)^2
\biggr\}, \\ 
 S_2^{(F)}=&\frac{1}{g_0^2}\sum_x \Tr \biggl\{
\Psi^{(0)T} G_\mu \frac{1}{2}\left(\Delta_\mu+\Delta_\mu^* \right) \Psi^{(0)}
+\Psi^{(0)T} P_\mu \frac{1}{2}\left(\Delta_\mu-\Delta_\mu^* \right)
 \Psi^{(0)}
+\Psi^{(0)T}   \cM \Psi^{(0)}
\biggr\}, 
\label{fermion kinetic}
\end{align}
respectively. 
Here $\Delta_\mu$ and $\Delta_\mu^*$ are forward and backward
difference operators; 
\begin{equation}
 \Delta_\mu f(x)=f(x+\hat\mu) - f(x), \quad 
 \Delta^*_\mu f(x)=f(x)-f(x-\hat\mu), 
\end{equation}
the matrices $G_\mu$ and $P_\mu$ are given by 
\begin{align}
 G_1 &=i\left(\begin{matrix}
&&&\sigma_1 \\
&&-i\sigma_2& \\
&i\sigma_2 && \\
\sigma_1 &&& 
\end{matrix}\right), \quad 
G_2 =i\left(\begin{matrix}
&&& -\sigma_3 \\
&&{\bf 1}_2& \\
&{\bf 1}_2 && \\
-\sigma_3 &&& 
\end{matrix}\right), \nn \\
P_1 &= i\left(\begin{matrix}
&&&i\sigma_2 \\
&&-\sigma_1& \\
&\sigma_1 && \\
i\sigma_2 &&& 
\end{matrix}\right), \quad 
P_2 = i\left(\begin{matrix}
&&&{\bf 1}_2 \\
&&\sigma_3& \\
&-\sigma_3 && \\
-{\bf 1}_2 &&& 
\end{matrix}\right),
\end{align}
and the mass matrix $\cM$ is 
\begin{align}
 \cM&=\left(\begin{matrix}
     &m_d \\
-m_d &
\end{matrix}\right), \quad 
m_d = {\rm
 diag}\left(\frac{M_0}{3},\frac{M_0}{3},\frac{M_0}{3}+\frac{m_0}{2},
-\frac{M_0}{3}\right). 
\end{align}
Note that $G_\mu$ and $P_\mu$ are anti-hermitian matrices and hermitian matrices, 
respectively, 
satisfying  
\begin{equation}
 \left\{G_\mu, G_\nu\right\}= -2 \delta_{\mu\nu}, \quad 
 \left\{P_\mu, P_\nu\right\}= 2 \delta_{\mu\nu}, \quad 
 \left\{G_\mu,P_\nu\right\}=0. 
\label{algPG}
\end{equation}

The bosonic part (\ref{boson kinetic}) takes the form of 
the standard lattice kinetic terms of bosons; 
no doubler appears in the bosonic sector. 
{}For the fermionic part (\ref{fermion kinetic}), 
the kernel in the momentum space takes the form, 
\begin{equation}
 \tilde{\cD}_F(p)=\sum_{\mu=1}^2\left[
iG_\mu \sin(a p_\mu) - 2P_\mu \sin^2\left(\frac{ap_\mu}{2}\right)
\right], 
\end{equation}
at $M_0=m_0=0$. 
Since the mass terms have the same structure as in the continuum, 
it is sufficient to consider the kinetic terms without the mass terms for our aim. 
Using (\ref{algPG}), we can easily see 
\begin{equation}
 \tilde{\cD}_F(p)^2 = \sum_{\mu=1}^2 4\sin^2\left(\frac{ap_\mu}{2}\right). 
\end{equation}
Since $\tilde{\cD}_F(p)$ is hermitian, it shows that only the origin 
$(p_1,p_2)=(0,0)$ gives the zero of $\tilde{\cD}_F(p)$ 
in $-\frac{\pi}{a}<p_\mu \leq \frac{\pi}{a}$, that is, 
there is no species doubler in the fermionic sector as well.

\subsection{Absence of parameter fine tunings}
Next, we discuss in the perturbation theory that the
desired quantum continuum theory 
is obtained without any fine tuning.
In the theory near the continuum limit with 
the auxiliary fields integrated out,
let us consider local operators of the type:
\begin{equation}
\cO_p(x)=\tilde{M}^{\tilde{m}}\varphi(x)^\alpha \del^\beta \psi(x)^{2\gamma}, 
\quad p \equiv \tilde{m} + \alpha + \beta + 3\gamma
\end{equation}
where $\varphi(x)$, $\psi(x)$ and $\del$ denote bosonic fields, 
fermionic fields and derivatives, respectively. 
$\tilde{M}$ represents $M$ or $m$. 
The mass dimension of $\cO_p$ is $p$ and 
$\tilde{m},\alpha,\beta,\gamma=0,1,2,\cdots$. 

{}From dimensional analysis, 
radiative corrections from ultraviolet (UV) 
region of loop momenta to $\cO_p$ have the form,  
\begin{equation}
\left(
\frac{1}{g_{2d}^2} c_0\, a^{p-4} + c_1\, a^{p-2} 
+ g_{2d}^2 c_2\, a^p + \cdots
\right)\int d^2 x \cO_p(x), 
\label{radiative correction}
\end{equation}
up to possible powers of $\ln(a\tilde{M})$. 
$c_0,c_1,c_2$ are dimensionless numerical constants.
The first, second and third terms in the parenthesis 
are contributions from tree,
1-loop and 2-loop effects, respectively.
The ``$\cdots$'' is an effect from higher loops, which
are irrelevant for the analysis.

Since relevant or marginal operators generated 
by loop effects possibly appear
from nonpositive powers of $a$ 
in the second and third terms in (\ref{radiative correction}), 
we should look at operators with $p=0,1,2$. 
They are $\varphi$, $\tilde{M}\varphi$ and $\varphi^2$ except for 
non-dynamical operators like $1$, $\tilde{M}$, $\tilde{M}^2$ and $\del\varphi$. 
For $G=U(N)$, although only the candidates for $\varphi$ is $\Tr B$ 
from gauge and $SU(2)_R$ symmetries, 
it is not invariant under $Q_\pm$ supersymmetries; it is forbidden to appear. 
Similarly, $\tilde{M}\varphi$ and $\varphi^2$ are not allowed to 
be radiatively generated. 
For $G=SU(N)$, we may consider $\varphi^2$ alone, 
whose candidates are not generated by the symmetries. 

Therefore, in the perturbative argument, we can conclude that any relevant
or marginal operators except non-dynamical operators do not appear radiatively,
meaning that no fine tuning is required to take the continuum limit.
In particular, if we consider the lattice theory around the trivial 
minimum $C=\phi_\pm=0$, the mass-deformed two-dimensional $\cN=(4,4)$ SYM is obtained 
without any fine-tuning. 
Also, after taking the limit of $m\to0$, 
we can safely take the limit of $M\to 0$ to reach the undeformed theory because of 
the exactly preserved eight supersymmetries. 
Thus we can use this lattice theory as a non-perturbative 
definition of two-dimensional $\cN=(4,4)$ SYM theory. 

\section{Four-dimensional $\cN=2$ supersymmetric Yang-Mills theory in the non-commutative space }

In this section, we discuss a scenario to obtain four-dimensional 
$\cN=2$ SYM on $\R^2\times ({\rm Fuzzy}\ \R^2)$ from the lattice formulation 
given in the previous section.

Let us consider the lattice theory for $G=U(N)$ around the minimum of $k$-coincident 
fuzzy sphere solution (\ref{FS solution}) with 
$M$ replaced by $M_0$, 
\begin{equation}
L_a=L_a^{(n)}\otimes {\bf 1}_k \quad (a=1,2,3) \quad \mbox{and}\quad N=nk.
\label{k_FS2}
\end{equation} 
$L_a^{(n)}$ are generators of an $n(=2j+1)$-dimensional irreducible representation of 
${\it su}(2)$ corresponding to spin $j$. 

{}First, we take the continuum limit of the two-dimensional lattice
theory. 
Then,
we obtain four-dimensional $\cN=2$ $U(k)$ SYM on 
$\R^2 \times ({\rm Fuzzy}\ S^2)$ deformed by the mass parameter $m$. 
The fuzzy $S^2$ has the radius $R=\frac{3}{M}$ 
and its noncommutativity is
characterized 
by the parameter $\Theta=\frac{18}{M^2n}$. 
The UV cutoff $\Lambda$ is naturally introduced by the size of the matrix; 
$\Lambda=\frac{M}{3}\cdot 2j$. 
Although these properties of the fuzzy $S^2$ can be seen by doing a similar calculation as presented in 
Refs.~\cite{Iso:2001mg,Maldacena:2002rb,Ishii:2008ib}, let us give a brief argument. 
Momentum modes of a field, say $B$, on two dimensions are expanded further by 
fuzzy spherical harmonics: 
\begin{equation}
\tilde{B}(q) = \sum_{J=0}^{2j}\sum_{m=-J}^J \hat{Y}^{(jj)}_{J\,m}\otimes b_{J\,m}(q),
\label{expand_B}
\end{equation}  
corresponding to the expression (\ref{k_FS2}).  
The fuzzy spherical harmonic $\hat{Y}^{(jj)}_{J\,m}$ is an $n\times n$ matrix whose elements are given by  
Clebsch-Gordon coefficients as~\cite{Ishii:2008ib}
\begin{equation}
\hat{Y}^{(jj)}_{J\,m} = \sqrt{n}\sum_{r,r'=-j}^j(-1)^{-j+r'}C^{J\,m}_{j\,r\,j\,-r'}\,|j\,r\rangle\langle j\,r'|
\end{equation}
with an orthonormal basis $|j\,r\rangle$ representing $L^{(n)}_a$ in the standard way: 
\begin{eqnarray}
\left(L_1^{(n)}\pm iL_2^{(n)}\right)\,|j\,r\rangle & = & \sqrt{(j\mp r)(j\pm r+1)}\,|j\,r\pm 1\rangle, \nn \\
L_3^{(n)}\,|j\,r\rangle & = & r\,|j\,r\rangle, 
\end{eqnarray}
and the modes $b_{J\,m}(q)$ are $k\times k$ matrices. 
It is seen that the fuzzy spherical harmonics are 
eigen-modes of the Laplacian on the fuzzy $S^2$: 
\be
\sum_{a=1}^3\left(\frac{M}{3}\right)^2[L^{(n)}_a, [L^{(n)}_a, \hat{Y}^{(jj)}_{J\,m}]] 
= \left(\frac{M}{3}\right)^2J(J+1)\hat{Y}^{(jj)}_{J\,m}, 
\ee
giving the rotational energy with the angular momentum $J$ on the sphere of the radius $R=\frac{3}{M}$.   
The UV cutoff $\Lambda= \frac{M}{3}\cdot 2j$ can be read off from the upper limit of the sum of $J$ 
in the expansion (\ref{expand_B}). 
The fuzzy $S^2$ is a two-dimensional non-commutative space, which is analogous to the phase space of 
some one-dimensional quantum system, 
and the noncommutativity $\Theta$ to the Planck constant $\hbar$. 
The quantum phase space is divided into small cells of the size $2\pi \hbar$, whose number is equal to 
the dimension of the Hilbert space. 
Correspondingly, the area of the $S^2$ is divided into $n$ cells of the size $2\pi\Theta$: 
\be
4\pi R^2 = n\cdot 2\pi \Theta,  
\ee
leading to the value $\Theta=\frac{18}{M^2n}$.    

As stressed in the previous section, 
the supersymmetry is softly broken from eight to two 
because of the mass parameter $m$ at this stage. 
The eight supercharges are recovered by taking the 
limit of $m\to 0$ with fixing $M$.

Next we take the limit of $n\to\infty$ with fixing $\Theta$ and $k$. 
In this limit, $M$ and $\Lambda$ become
\begin{equation}
M\propto n^{-1/2}\to 0, \quad 
\Lambda\propto n^{1/2}\to \infty, 
\end{equation}
and the fuzzy $S^2$ is decompactified to the non-commutative 
Moyal plane $\R_\Theta^2$. 
Since the fuzzy $S^2$ solution preserves eight supercharges
after taking $m\to0$, 
it is strongly expected that the theory becomes $\cN=2$ $U(k)$ 
SYM on $\R^2\times \R_\Theta^2$ after taking the above limit. 
The gauge coupling constant of the four-dimensional theory is given in the form 
\begin{equation}
g_{4d}^2 =2\pi \Theta g_{2d}^2. 
\label{4d_2d_coupling}
\end{equation} 
After taking this limit, 
the expansion (\ref{expand_B}) by the fuzzy spherical harmonics can be 
essentially transcribed to  
the one by plane waves on ${\R}^2_{\Theta}$:  
\be
\tilde{B}(q) = \int\frac{d^2\tilde{q}}{(2\pi)^2}\,e^{i\tilde{q}\cdot\hat{x}}\otimes \tilde{b}({\bf q}), 
\ee
where $q$ and $\tilde{q}$ are two-momenta on $\R^2$ and $\R^2_{\Theta}$ respectively,  
the position operator $\hat{x}=(\hat{x}_1, \hat{x}_2)$ on $\R^2_{\Theta}$ satisfies 
$[\hat{x}_1, \hat{x}_2]=i\Theta$, and ${\bf q}\equiv (q,\tilde{q})$ represents a four-momentum. 
The modes $\tilde{b}({\bf q})$ in the four-dimensional space are $k\times k$ matrices.  
It is easy to calculate the inner product between plane waves on $\R^2_{\Theta}$:   
\be
\Tr\left(e^{i\tilde{p}\cdot\hat{x}}e^{i\tilde{q}\cdot\hat{x}}\right) 
= \frac{2\pi}{\Theta}\,\delta^2(\tilde{p}+\tilde{q}),  
\ee
which leads to the $\Theta$-dependence of the relation (\ref{4d_2d_coupling}). 

Although more investigation is needed to clarify whether 
the $\Theta\to 0$ limit of the theory is 
continuously connected to the commutative four-dimensional 
$\cN=2$ SYM on $\R^4$ or not%
\footnote{
It is naively expected that the $\Theta\to 0$ limit 
would not be continuously connected to the commutative theory 
because of the ultraviolet/infrared (UV/IR) mixing \cite{Minwalla:1999px}. 
There is a discussion, however, 
that non-commutative four-dimensional $\cN=2$ $U(k)$ SYM may flow to 
the ordinary commutative theory in the infrared
\cite{Armoni:2001br}. 
}, 
to the best of our knowledge 
this formulation gives the first non-perturbative formulation free from fine tuning for 
four-dimensional SYM with eight supercharges.

\section{Conclusion and discussion}
In this paper, 
we deformed two-dimensional $\cN=(4,4)$ SYM theory 
with the gauge group $U(N)$ or $SU(N)$
by a mass parameter $M$ with preserving all supercharges 
and expressed the deformed action in BTFT form. 
We further deformed it by introducing an additional 
mass parameter $m$ in a manner to keep two supercharges, $Q_\pm$,  
in order to lift up all the flat directions of the scalar fields. 
We then put the deformed theory on a two-dimensional lattice 
with preserving $Q_\pm$ exactly.
The problem of 
the running of the vacuum expectation values of the scalar fields 
is avoided thanks to the deformation by $M$ and $m$. 
We also gave a perturbative argument that any fine tuning is not needed 
in taking the lattice continuum limit. 
Thus this lattice theory around the trivial minimum 
can be regarded 
as a non-perturbative definition of two-dimensional $\cN=(4,4)$ 
SYM theory. 
To perform actual numerical simulation, it should be checked if the
imaginary term of
the bosonic action (\ref{boson part}) is managed by the reweighting
method, 
which might cause bosonic sign problem independent of the fermionic
one. 
We next considered the lattice theory for the gauge group $U(N)$ around a $k$-coincident 
fuzzy sphere solution with $N=nk$. 
The radius of the fuzzy $S^2$ is $\frac{3}{M}$ and the noncommutativity 
of the fuzzy sphere is characterized by the parameter $\Theta=\frac{18}{M^2n}$. 
By taking the lattice continuum limit, we obtained four-dimensional 
$\cN=2$ $U(k)$ SYM on $\R^2\times({\rm Fuzzy}\ S^2)$ deformed by 
the mass parameter $m$. 
It was discussed that, by taking the limit of 
$m\to 0$ followed by the limit of $M\to 0$ with fixing $k$ and $\Theta$, 
four-dimensional $\cN=2$ SYM on $\R^2\times \R^2_\Theta$ is
realized without any fine tuning. 

In contrast to four-dimensional $\cN=4$ SYM, 
the commutative limit $\Theta\to 0$ of 
four-dimensional $\cN=2$ non-commutative SYM 
is expected not to be continuously connected to 
the usual commutative $\cN=2$ SYM because of UV/IR mixing. 
Even if such expectation is true and our scenario does not lead to $\cN=2$ SYM on the usual $\R^4$, 
notice that 
non-commutative gauge theory itself is an important subject of
research in order to clarify non-perturbative aspects of 
gauge theories. 
{}For example, when we consider instantons of gauge theories, 
noncommutativity plays a crucial role to resolve 
the small instanton singularity.
In our formulation, we can analyze the dynamical aspect of 
quite wide class of observables 
of four-dimensional $\cN=2$ non-commutative SYM numerically,  
which will give a strong instrument to reveal the non-perturbative 
structure of supersymmetric gauge theories.

The deformed two-dimensional theory itself is also interesting on its own. 
In particular, since one can introduce mass terms for all scalars 
keeping two supersymmetries, and hence flat directions (along which 
all scalars commute each other) are all lifted, one can perform stable Monte-Carlo 
simulation, if reweighting for the imaginary term in (\ref{boson part}) works. 
(Simulations so far utilized supersymmetry breaking mass terms, which make 
the conclusion more or less obscure.)  
Moreover, the deformation terms consist of mass terms and a Myers
term, which are quite similar to the so-called $\Omega$-deformation 
\cite{Nekrasov:2003af,Nekrasov:2003rj}
which is originally introduced in order to regularize
the instanton moduli space of four-dimensional $\cN=2$ SYM theory%
\footnote{ 
In fact, starting with one-dimensional theory that is obtained 
by the dimensional reduction of four-dimensional $\cN=2$ SYM 
in the $\Omega$-background, 
we can construct a regularized three-dimensional theory on 
$\R\times ({\rm Fuzzy}\ S^2)$ 
with keeping at least a part of the supersymmetry \cite{unpublished}.}.
In the case of the $\Omega$-deformation, the integration 
over the instanton moduli space is localized to discrete points, 
which makes it possible to evaluate the instanton partition function 
analytically using localization formula in equivariant cohomology. 
On the other hand, the deformation introduced in this paper 
lifts flat directions of the scalar fields, 
which makes it possible to carry out stable Monte-Carlo simulation. 
It is interesting that a mathematically sophisticated technique 
like equivariant cohomology somehow relates to 
a technique 
developed for numerical simulation in this paper. 
It may be a sign that such a mathematical method 
would give a systematic method to construct a non-perturbative 
definition of supersymmetric gauge theories in the future.

\section*{Acknowledgments}
The authors would like to thank Adi Armoni, 
Hikaru Kawai, Noboru Kawamoto, Jun Nishimura, 
Hidehiko Shimada, 
Hiroshi Suzuki, Asato Tsuchiya and 
Mithat \"{U}nsal for  
discussions and enlightening comments. 
M.~H. and F.~S. would like to thank Weizmann Institute for Science 
where this work was initiated.      
The work of M.~H. is supported from Postdoctoral Fellowship for Research Abroad 
by Japan Society for the Promotion of Science.
The work of S.~M.~ is supported in part by Grant-in-Aid
for Young Scientists (B), 23740197
and Keio Gijuku Academic Development Funds. 
The work of F.~S.~ is supported in part by Grant-in-Aid
for Scientific Research (C), 21540290. 

\appendix 

\section{Plane wave deformed two-dimensional ${\cal N}=(4,4)$ supersymmetric Yang-Mills theory} 
\label{app.A}

In this appendix we explain how to construct the plane wave deformed two-dimensional ${\cal N}=(4,4)$ supersymmetric Yang-Mills theory. 
\subsection{BMN type matrix model with 8 supercharges} 
Let us start with an eight-supersymmetry analogue \cite{Kim:2006wg} of the plane wave matrix
model \cite{Berenstein:2002jq}, 
\bea
S & = & R\int dt \,\Tr\left[\frac{1}{2R^2}(D_tX^i)^2 + \frac{i}{R}\Psi^TD_t\Psi 
+\Psi^T\Gamma^i[X^i, \Psi]+\frac14[X^i,X^j]^2 \right.\nn \\
 & & \left.-\frac12\left(\frac{\mu}{3R}\right)^2 (X^a)^2 -\frac12\left(\frac{\mu}{6R}\right)^2(X^{a'})^2 
 -i\frac{\mu}{4R}\Psi^T\Gamma^{456}\Psi -i\frac{\mu}{3R}\epsilon_{abc}X^aX^bX^c\right], \nn \\
& &  
\label{S_6DBMN}
\eea
where $D_t = \del_t +i[A_t,\ \cdot\ ]$, $i=2,3,4,5,6$, $a=4,5,6$ and $a'=2,3$. 
$\Gamma^i$ are $8\times 8$ real symmetric matrices corresponding to
$-i\gamma_i$, 
which satisfy 
\bea
 & & \{\Gamma^i, \Gamma^j\} = 2\delta^{ij}{\bf 1}_8,  \\
 & & \Gamma^{23456} = \Gamma^2\cdots\Gamma^6 = -1.
\eea
 
{}From this model we construct a two-dimensional theory following \cite{Das:2003yq},  
by using Taylor's T-duality. 
In order to lift the theory to two dimensions, 
we redefine the fields by a rotation on $(X^2, X^3)$ plane with the angle $\alpha t$ as 
\be
 \begin{pmatrix} X^2 \\ X^3 \end{pmatrix} = 
U_\alpha(t) \begin{pmatrix} \hat{X}^2 \\ \hat{X}^3 \end{pmatrix} , \qquad 
\Psi = e^{\frac12 \Gamma^{23}\alpha t}\hat{\Psi}
\ee
with 
\be
U_\alpha (t)\equiv \begin{pmatrix} \cos(\alpha t) & \sin(\alpha t) \\ -\sin(\alpha t) & \cos(\alpha t) 
\end{pmatrix}. 
\ee
{}For the other variables, the hatted variables are the same as the unhatted ones.  
{}For example, $D_t = \del_t +i[A_t, \cdot]=\del_t +i[\hat{A}_t, \cdot]$.
Since $\Gamma^{23}$ is real anti-symmetric, $\Psi^T$ transforms as 
$\Psi^T = \hat{\Psi}^Te^{-\frac12 \Gamma^{23}\alpha t}$. 
The action in terms of the redefined fields is 
\bea
S & = & R\int dt \,\Tr \left[\frac{1}{2R^2}(D_t\hat{X}^i)^2+ \frac{i}{R}\hat{\Psi}^TD_t\hat{\Psi} 
+\hat{\Psi}^T\Gamma^i[\hat{X}^i, \hat{\Psi}]+\frac14[\hat{X}^i,\hat{X}^j]^2 \right.\nn \\
 & & 
-\frac12\left\{\left(\frac{\mu}{6R}\right)^2-\frac{\alpha^2}{R^2}\right\} (\hat{X}^{a'})^2 
 -i\frac{\mu}{4R}\hat{\Psi}^T\left(\Gamma^{456}-\frac{2\alpha}{\mu}\Gamma^{23}\right)\hat{\Psi} 
\nn \\
 & & \left.+\frac{2\alpha}{R^2}\hat{X}^3D_t\hat{X}^2 -\frac12\left(\frac{\mu}{3R}\right)^2 (\hat{X}^a)^2  -i\frac{\mu}{3R}\epsilon_{abc}\hat{X}^a\hat{X}^b\hat{X}^c\right] \nn \\
 & & -\frac{\alpha}{R}\int \,dt \,\del_t\Tr\, (\hat{X}^2\hat{X}^3).  
\label{S_6DBMN_hat}
\eea
If we set $\alpha =\pm \frac{\mu}{6}$ and discard the surface term, 
the mass terms of $\hat{X}^{a'}$ vanish, and $\hat{X}^2$ appears only in the adjoint form. 
Then, the Taylor's T-duality may be performed with respect 
to $\hat{X}^2$. 
As we will see shortly, compatibility of supersymmetry transformation and T-duality 
singles out $\alpha=-\frac{\mu}{6}$. 
\subsubsection{Supersymmetry transformation} 
Supersymmetry transformation of this matrix quantum mechanics is given by 
\bea
\delta X^i & = & i\Psi^T\Gamma^i\epsilon(t), \nn \\
\delta A_t & = & Ri\Psi^T\epsilon(t), \nn \\
\delta \Psi & = & \left\{\frac{1}{2R}(D_t X^i) \Gamma^i +\frac{i}{4} [X^i,X^j] \Gamma^{ij} \right. \nn \\ 
 & & \left. +\frac{\mu}{6R} X^a \Gamma^a \Gamma^{456} -\frac{\mu}{12R} X^{a'}\Gamma^{a'}\Gamma^{456} 
 \right\} \epsilon(t)
\eea
with 
\be
\epsilon(t) = e^{-\frac{\mu}{12}\Gamma^{456}t}\epsilon_0, 
\ee
where $\epsilon_0$ is an 8-component constant spinor. 
This is called ``dynamical supersymmetry'' which is simply referred as supersymmetry in the text. 
{}For the case of $G=U(N)$, it is also invariant under ``kinematical supersymmetry'',   
\be
\tdelta \Psi=\eta(t) {\bf 1}_N
\ee
with 
\be
\eta(t) = e^{\frac{\mu}{4}\Gamma^{456}t}\eta_0. 
\ee
 
In terms of the redefined fields, the dynamical supersymmetry transformation becomes  
\bea
\delta \hat{X}^i & = & i\hat{\Psi}^T\Gamma^i \hat{\epsilon}(t), \nn \\
\delta \hat{A}_t & = & Ri\hat{\Psi}^T\hat{\epsilon}(t), \nn \\
\delta \hat{\Psi} & = & \left\{\frac{1}{2R}(D_t\hat{X}^i)\Gamma^i +\frac{i}{4}[\hat{X}^i,\hat{X}^j]\Gamma^{ij} 
+\frac{\mu}{6R}\hat{X}^a\Gamma^a \Gamma^{456} \right. \nn \\
 & & \left. -\frac{1}{2R}\left(\frac{\mu}{6}+\alpha\right)\left(\hat{X}^2\Gamma^3-\hat{X}^3\Gamma^2\right) 
\right\} \hat{\epsilon}(t), 
\eea
where $\hat{\epsilon}$ is given by $\epsilon(t)=e^{\frac12 \Gamma^{23}\alpha t} \hat{\epsilon}(t)$. 
Here we used $\Gamma^{456}=\Gamma^{23}$ to obtain 
\be
\hat{X}^{a'}\Gamma^{a'}\Gamma^{456} = \left(\hat{X}^2\Gamma^2 + \hat{X}^3\Gamma^3\right) \Gamma^{23} 
= \hat{X}^2\Gamma^3 -\hat{X}^3\Gamma^2.
\ee
{}For the case that $\hat{X}^2$ appears only in the adjoint form in the supersymmetry transformation, 
Taylor's T-duality procedure keeps the supersymmetry. 
It uniquely fixes the choice of $\alpha$ to 
\be
\alpha = -\frac{\mu}{6}.   
\ee
Then, the parameter $\hat{\epsilon}(t)$ becomes $t$-independent: 
\be
\hat{\epsilon}(t) = e^{-\frac12\Gamma^{23}\alpha t}\epsilon(t) 
= e^{\frac{\mu}{12}(\Gamma^{23}-\Gamma^{456})t}\epsilon_0 = \epsilon_0,  
\ee
and the dynamical supersymmetry is expressed as  
\bea
\delta\hat{X}^i & = & i\hat{\Psi}^T\Gamma^i\epsilon_0, \nn \\
\delta\hat{A}_t & = & Ri\hat{\Psi}^T\epsilon_0, \nn \\
\delta \hat{\Psi} & = & \left\{\frac{1}{2R}(D_t\hat{X}^i)\Gamma^i +\frac{i}{4}[\hat{X}^i,\hat{X}^j]\Gamma^{ij} 
+\frac{\mu}{6R}\hat{X}^a\Gamma^a \Gamma^{456} \right\}\epsilon_0.
\eea

Also, the kinematical supersymmetry becomes 
\be
\tdelta \hat{\Psi} = \hat{\eta}(t){\bf 1}_N 
\ee
with 
\be
\hat{\eta}(t) =   e^{-\frac12 \Gamma^{23}\alpha t} \eta(t) 
=e^{\frac{\mu}{4}\left(\Gamma^{456}+\frac13\Gamma^{23}\right) t}\eta_0 
 = e^{\frac{\mu}{3}\Gamma^{23}t}\eta_0.
\ee
 
The final form of the action (\ref{S_6DBMN_hat}) with the surface term dropped is 
\bea
S & = & R\int dt \,\Tr \left[\frac{1}{2R^2}(D_t\hat{X}^i)^2+ \frac{i}{R}\hat{\Psi}^TD_t\hat{\Psi} 
+\hat{\Psi}^T\Gamma^i[\hat{X}^i, \hat{\Psi}]+\frac14[\hat{X}^i,\hat{X}^j]^2 \right.\nn \\
 & & \left.
 -i\frac{\mu}{3R}\hat{\Psi}^T\Gamma^{23}\hat{\Psi} 
-\frac{\mu}{3R^2}\hat{X}^3D_t\hat{X}^2 -\frac12\left(\frac{\mu}{3R}\right)^2 (\hat{X}^a)^2
 -i\frac{\mu}{3R}\epsilon_{abc}\hat{X}^a\hat{X}^b\hat{X}^c\right].\nn \\
 & & \label{S_6DBMN_f}
\eea
\subsection{Uplift to two dimensions: Taylor's T-duality}
In order to obtain a two-dimensional theory, 
we ``compactify'' the $\hat{X}^2$ direction to a circle of a radius $\hat{R}$ as 
\be
\hat{X}^2 \sim \hat{X}^2+2\pi \hat{R}.
\ee
By using the argument by Taylor \cite{Taylor:1996ik}, 
the action (\ref{S_6DBMN_f}) is lifted to two dimensions,  
\bea
S & = & R\hat{R}\int dt \int^{1/\hat{R}}_0d\sigma \,\Tr\left[\frac{1}{2R^2}F_{t\sigma}^2 
+\frac{1}{2R^2}(D_tX^I)^2 -\frac12 (D_\sigma X^I)^2 \right. \nn \\
 & & +\frac{i}{R}\Psi^TD_t\Psi -i\Psi^T\Gamma^2 D_\sigma \Psi  
+\Psi^T\Gamma^I[X^I,\Psi] +\frac14 [X^I, X^J]^2 \nn \\
 & & \left. -\frac12\left(\frac{\mu}{3R}\right)^2(X^a)^2 
-i\frac{\mu}{3R}\Psi^T\Gamma^{23}\Psi 
-\frac{\mu}{3R^2}X^3 F_{t\sigma} -i\frac{\mu}{3R}\epsilon_{abc}X^aX^bX^c \right], \nn \\
 & & 
\eea 
where $I=3,4,5,6$. 
(The hats of the fields were omitted.) 

By setting 
\begin{align}
&t=\frac{1}{R}x_1, \quad A_t = RA_1, \quad \sigma =x_2, \nn \\ 
&1/\hat{R}=L, \quad g=(2L)^{1/2}, \quad M=\frac{\mu}{R}, 
\end{align}
and by rescaling the fermion as $\Psi \to \frac{1}{\sqrt{2}} \Psi$, 
we obtain 
\bea
S & = & \frac{2}{g^2}\int d^2x \,\Tr \left[\frac12 F_{12}^2 +\frac12 (D_1X^I)^2 -\frac12 (D_2 X^I)^2 \right. 
\nn \\ 
 & & +\frac{i}{2}\Psi^T\left(D_1-\Gamma^2 D_2\right) \Psi +\frac12 \Psi^T\Gamma^I[X^I,\Psi] 
 +\frac14 [X^I, X^J]^2 \nn \\
 & & \left. -\frac12 \left(\frac{M}{3}\right)^2(X^a)^2 -i\frac{M}{6}\Psi^T\Gamma^{23}\Psi 
 -\frac{M}{3}X^3F_{12} -i\frac{M}{3}\epsilon_{abc}X^a X^b X^c \right], \nn \\
 & & 
\eea
where 
\be
\int d^2x \cdots \equiv \int dx_1 \int^L_0 dx_2 \cdots. 
\ee

The dynamical supersymmetry is expressed as 
\bea
\delta A_1 & = & -i\Psi^T\epsilon, \nn \\
\delta A_2 & = & -i\Psi^T\Gamma^2\epsilon, \nn \\
\delta X^I & = &- i\Psi^T \Gamma^I \epsilon, \nn \\
\delta \Psi & = & -\left\{ \frac{}{}F_{12} \Gamma^2 +(D_1 X^I)\Gamma^I + (D_2 X^I)\Gamma^{2I} \right. \nn \\
 & & \left. +\frac{i}{2}[X^I, X^J]\Gamma^{IJ} +\frac{M}{3} X^a \Gamma^a\Gamma^{456}\right\} \epsilon, 
\eea 
where $\epsilon=-\epsilon_0 /\sqrt{2}$ is an 8-component constant spinor. 
Note that all the dynamical supersymmetries are preserved. It is in sharp contrast with 
the case of sixteen supercharges \cite{Das:2003yq}, 
where a half of supersymmetries are broken by deformations.
 
The kinematical supersymmetry also remains. It is given by  
\be
\tdelta \Psi = \hat{\eta}(x_1){\bf 1}_N
\ee
with 
\be
\hat{\eta}(x_1) = e^{\frac{M}{3}\Gamma^{23}x_1}\eta_0 , 
\ee
where $\eta_0$ is constant. 
\subsubsection{Wick rotation}   
In order to obtain the Euclidean action, which is going to be put on a lattice, 
we perform the Wick rotation, 
\be
x_1 \to -i x_1, \qquad A_1 \to iA_1.
\ee
The Euclidean action is 
\bea
S_E & = & \frac{2}{g^2}\int d^2x \,  \Tr \left[ \frac12 F_{12}^2 +\frac12 (D_\mu X^I)^2 
 +\frac12\Psi^T\left(D_1+\gamma_2 D_2\right) \Psi \right. \nn \\
 & & +\frac{i}{2} \Psi^T\gamma_I [X^I,\Psi] -\frac14 [X^I, X^J]^2 \nn \\
 & & \left. +\frac12 \left(\frac{M}{3}\right)^2(X^a)^2 -i\frac{M}{6}\Psi^T\gamma_{23}\Psi 
 +i\frac{M}{3}X^3F_{12} +i\frac{M}{3}\epsilon_{abc}X^a X^b X^c \right], \nn \\
 & & 
\label{SE_6DBMN} 
\eea
where $\mu =1,2$, and $\gamma_I = i\Gamma^I$, 
which is identical to (\ref{2dSYM-M}). 

Then, the dynamical supersymmetry transformation is written as 
\bea
\delta A_1 & = & \epsilon^T \Psi, \nn \\
\delta A_2 & = & \epsilon^T \gamma_2\Psi, \nn \\
\delta X^I  & = & \epsilon^T \gamma_I\Psi, \nn \\
\delta \Psi & = & \left(-F_{12}\gamma_2 -(D_1X^I)\gamma_I +(D_2 X^I)\gamma_{2I} 
+\frac{i}{2} [X^I, X^J]\gamma_{IJ} -\frac{M}{3} X^a\gamma_a
 \gamma_{456}\right)  \epsilon, \nn \\
\label{dyn_SUSY_2dE} 
\eea
and the kinematical supersymmetry is given by 
\be
\delta'\Psi = \hat{\eta}(x_1){\bf 1}_N
\label{kin_SUSY_2dE}
\ee
with 
\be
\hat{\eta}(x_1) = e^{i\frac{M}{3}\gamma_{23}x_1}\eta_0.
\ee

\section{Explicit form of the lattice action}
\label{app.B}

In this section, we explicitly write down the lattice action 
(\ref{lattice action}) in terms of lattice fields.  

We divide the action into the bosonic and the fermionic parts; 
\begin{equation}
 S_{\rm lat} = S_{\rm lat}^{(B)} + S_{\rm lat}^{(F)}. 
\end{equation}
The bosonic part is given by
\begin{align}
 S_{\rm lat}^{(B)}=&\frac{1}{g_0^2}\sum_x \Tr \Biggl[
H(x)^2 -2i H(x) A(x) 
+\tH_\mu(x)^2+2i \tH_\mu(x) \tA_\mu(x) \nn \\
&+\left(\cD_\mu \phi_+(x)\right)\left(\cD_\mu \phi_-(x)\right)
+\frac{1}{4}\left(\cD_\mu C(x)\right)^2 \nn \\
&-\frac{1}{4}\left[B(x),C(x)\right]^2 
-\left[B(x),\phi_+(x)\right]\left[B(x),\phi_-(x)\right] \nn \\
&+\frac{1}{4}\left[\phi_+(x),\phi_-(x)\right]^2
-\frac{1}{4}\left[C(x),\phi_+(x)\right]\left[C(x),\phi_-(x)\right] \nn \\
&-\frac{M_0}{2}C(x)\left[\phi_+(x),\phi_-(x)\right]
+\frac{M_0^2}{9}\left(\frac{1}{4}C(x)^2+\phi_+(x)\phi_-(x)\right) \nn \\
&-\frac{M_0m_0}{6} B(x)^2 + i \frac{M_0}{3}B(x)\hPhi(x) 
\Biggr], 
\label{lattice action bosonic pre}
\end{align}
where 
\begin{align}
 A(x)=&\frac{1}{2}\hPhi(x)+\frac{i}{2}m_0 B(x), \nn \\
 \tA_1(x)=& \frac{1}{2}\frac{1}{1-\frac{1}{\e^2}||1-U_{12}(x)||^2}
 \nn \\
&\times \biggl[
-U_{12}(x)B(x)-B(x)U_{21}(x) \nn \\
&\qquad +U_2(x-\hat{2})^{-1}\left(
B(x-\hat{2})U_{12}(x-\hat{2})+U_{21}(x-\hat{2})B(x-\hat{2})
\right)U_2(x-\hat{2}) 
\biggr] \nn \\
+&\frac{1}{2\e^2}
\frac{\Tr\Bigl(B(x)\left(U_{12}(x)-U_{21}(x)\right)\Bigr)}
{\left(1-\frac{1}{\e^2}||1-U_{12}(x)||^2\right)^2}
\Bigl(U_{12}(x)-U_{21}(x)\Bigr) \nn \\
-&\frac{1}{2\e^2}
\frac{\Tr\Bigl(B(x-\hat{2})\left(U_{12}(x-\hat{2})-U_{21}(x-\hat{2})\right)\Bigr)}
{\left(1-\frac{1}{\e^2}||1-U_{12}(x-\hat{2})||^2\right)^2} \nn \\
&\qquad \times \Bigl(U_2(x-\hat{2})^{-1}\left(
U_{12}(x-\hat{2})-U_{21}(x-\hat{2})
\right)U_2(x-\hat{2})\Bigr), \nn \\
 \tA_2(x)=& \frac{1}{2}\frac{1}{1-\frac{1}{\e^2}||1-U_{12}(x)||^2}
 \nn \\
&\times \biggl[
B(x)U_{12}(x)+U_{21}(x)B(x) \nn \\
&\qquad -U_1(x-\hat{1})^{-1}\left(
U_{12}(x-\hat{1})B(x-\hat{1})+B(x-\hat{1})U_{21}(x-\hat{1})
\right)U_1(x-\hat{1}) 
\biggr] \nn \\
-&\frac{1}{2\e^2}
\frac{\Tr\Bigl(B(x)\left(U_{12}(x)-U_{21}(x)\right)\Bigr)}
{\left(1-\frac{1}{\e^2}||1-U_{12}(x)||^2\right)^2}
\Bigl(U_{12}(x)-U_{21}(x)\Bigr) \nn \\
+&\frac{1}{2\e^2}
\frac{\Tr\Bigl(B(x-\hat{1})\left(U_{12}(x-\hat{1})-U_{21}(x-\hat{1})\right)\Bigr)}
{\left(1-\frac{1}{\e^2}||1-U_{12}(x-\hat{1})||^2\right)^2} \nn \\
&\qquad \times \Bigl(U_1(x-\hat{1})^{-1}\left(
U_{12}(x-\hat{1})-U_{21}(x-\hat{1})
\right)U_1(x-\hat{1})\Bigr). 
\end{align}
Note that $\tA_\mu(x)$ are hermitian. 
After integrating out the auxiliary fields, $S_{\rm lat}^{(B)}$ becomes 
\begin{equation}
 S_{\rm lat}^{(B)}=\frac{1}{g_0^2}\sum_x \Tr \Biggl[ 
-\frac{m_0}{2}\left(\frac{M_0}{3}+\frac{m_0}{2}\right) B(x)^2 + 
i\left( \frac{M_0}{3}+\frac{m_0}{2}\right)B(x)\hPhi(x) 
\Biggr]
+S_{\rm PDT}, 
\label{lattice action bosonic}
\end{equation}
where $S_{\rm PDT}$ denotes positive (semi-)definite terms: 
\begin{align}
 S_{\rm PDT} = &\frac{1}{g_0^2}\sum_x \Tr \Biggl[ 
\frac{1}{4}\hPhi(x)^2+\left(\cD_\mu \phi_+(x)\right)\left(\cD_\mu \phi_-(x)\right)
+\frac{1}{4}\left(\cD_\mu C(x)\right)^2 \nn \\
&-\frac{1}{4}\left[B(x),C(x)\right]^2 
-\left[B(x),\phi_+(x)\right]\left[B(x),\phi_-(x)\right] \nn \\
&+\frac{1}{4}\left[\phi_+(x),\phi_-(x)\right]^2
-\frac{1}{4}\left[C(x),\phi_+(x)\right]\left[C(x),\phi_-(x)\right] \nn \\
&-\frac{M_0}{2}C(x)\left[\phi_+(x),\phi_-(x)\right]
+\frac{M_0^2}{9}\left(\frac{1}{4}C(x)^2+\phi_+(x)\phi_-(x)\right) +\tA_1(x)^2+\tA_2(x)^2
\Biggr]. 
\label{PDT}
\end{align}
In order that the field $B(x)$ has positive mass squared, 
$m_0$ must satisfy 
 \begin{equation}
 -\frac{2M_0}{3} < m_0 < 0. 
\end{equation}

The fermionic part is given by
\begin{align}
 S_{\rm lat}^{(F)}=&\frac{1}{g_0^2}\sum_x \Tr \Biggl[
i\psi_{+\mu}(x)\cD_\mu \eta_-(x)
+i\psi_{-\mu}(x)\cD_\mu\eta_+(x) \nn \\
&+\chi_+(x)[\phi_-(x),\chi_+(x)]
-\chi_-(x)[\phi_+(x),\chi_-(x)]
+\chi_+(x)[C(x),\chi_-(x)] \nn \\
&-\eta_+(x)[B(x),\chi_-(x)]
-\eta_-(x)[B(x),\chi_+(x)] \nn \\
&+\frac{1}{4}\eta_+(x)[\phi_-(x),\eta_+(x)]
-\frac{1}{4}\eta_-(x)[\phi_+(x),\eta_-(x)]
-\frac{1}{4}\eta_+(x)[C(x),\eta_-(x)] \nn \\
&-\psi_{+\mu}(x)\psi_{+\mu}(x)\Bigl(\phi_-(x)+U_\mu(x)\phi_-(x+\hmu)U_\mu(x)^{-1}\Bigr)
 \nn \\
&+\psi_{-\mu}(x)\psi_{-\mu}(x)\Bigl(\phi_+(x)+U_\mu(x)\phi_+(x+\hmu)U_\mu(x)^{-1}\Bigr)
 \nn \\
&-\frac{1}{2}\left\{\psi_{+\mu}(x),\psi_{-\mu}(x)\right\}
\Bigl(C(x)+U_\mu(x) C(x+\hmu)U_\mu(x)^{-1}\Bigr) \nn \\
&+\frac{1}{2}\psi_{+\mu}(x)\psi_{+\mu}(x)\psi_{-\mu}(x)\psi_{-\mu}(x)
 \nn \\
&+\frac{2M_0}{3}\psi_{+\mu}(x)\psi_{-\mu}(x)
+\left(\frac{2M_0}{3}+m_0\right)\chi_+(x)\chi_-(x)
-\frac{M_0}{6}\eta_+(x)\eta_-(x) \nn \\
&+i\chi_-(x)\Bigl(Q_+ \hPhi(x)\Bigr)
-i\chi_+(x)\Bigl(Q_-\hPhi(x)\Bigr)
-iB(x)\Bigl(Q_+ Q_- \hPhi(x)\Bigr)\Bigr|_{\rm fermion}
\Biggr], 
\end{align}
where 
\begin{align}
 Q_\pm\hPhi(x)=&\frac{-1}{1-\frac{1}{\e^2}||1-U_{12}(x)||^2} \nn \\
&\times\biggl[
-\left(\psi_{\pm1}(x)+U_1(x)\psi_{\pm2}(x+\hat{1})U_1(x)^{-1}\right)U_{12}(x)
 \nn \\
&\qquad
-U_{21}(x)\left(\psi_{\pm1}(x)+U_1(x)\psi_{\pm2}(x+\hat{1})U_1(x)^{-1}\right)
 \nn \\
&\qquad
+\left(\psi_{\pm2}(x)+U_2(x)\psi_{\pm1}(x+\hat{2})U_2(x)^{-1}\right)U_{21}(x)
 \nn \\
&\qquad
+U_{12}(x)\left(\psi_{\pm2}(x)+U_2(x)\psi_{\pm1}(x+\hat{2})U_2(x)^{-1}\right)
\biggr] \nn \\
&+\frac{U_{12}(x)-U_{21}(x)}{\left(1-\frac{1}{\e^2}||1-U_{12}(x)||^2\right)^2}
\frac{1}{\e^2} \nn \\
&\times\Tr\biggl[
\Bigl(U_{12}(x)-U_{21}(x)\Bigr)
\Bigl(
\cD_2 \psi_{\pm1}(x)-\cD_1 \psi_{\pm2}(x)
\Bigr)
\biggr],  \\
Q_+Q_-\hPhi(x)\Bigr|_{\rm fermion}=&
 -i\frac{Q_+Q_-\left(U_{12}(x)-U_{21}(x)\right)\Bigr|_{\rm fermion}}
{1-\frac{1}{\e^2}||1-U_{12}(x)||^2} \nn \\
&+\frac{i\left(U_{12}(x)-U_{21}(x)\right)}
{\Bigl(1-\frac{1}{\e^2}||1-U_{12}(x)||^2\Bigr)^2}
\frac{1}{\e^2}\Tr\Bigl[
Q_+Q_-\left(U_{12}(x)+U_{21}(x)\right)
\Bigr]\Bigr|_{\rm fermion} \nn \\
&-\frac{1}{\Bigl(1-\frac{1}{\e^2}||1-U_{12}(x)||^2\Bigr)^2}
\frac{1}{\e^2} \nn \\
&\quad\times \Bigl\{
-iQ_+\left(U_{12}(x)-U_{21}(x)\right)
\Tr\Bigl[Q_-\left(U_{12}(x)+U_{21}(x)\right)\Bigr] \nn \\
&\quad\qquad
+iQ_-\left(U_{12}(x)-U_{21}(x)\right)
\Tr\Bigl[Q_+\left(U_{12}(x)+U_{21}(x)\right)\Bigr]
\Bigr\} \nn \\
&-2\frac{i\left(U_{12}(x)-U_{21}(x)\right)}
{\Bigl(1-\frac{1}{\e^2}||1-U_{12}(x)||^2\Bigr)^3}
\frac{1}{\e^4} \nn \\
&\quad\times \Tr\Bigl[Q_+\left(U_{12}(x)+U_{21}(x)\right)\Bigr]
\Tr\Bigl[Q_-\left(U_{12}(x)+U_{21}(x)\right)\Bigr], 
\end{align}
with 
\begin{align}
&Q_+Q_- \left(U_{12}(x)-U_{21}(x)\right)\Bigr|_{\rm fermion} \nn \\
&\qquad = 
\Bigl\{-\frac{1}{2}\left[\psi_{+1}(x),\psi_{-1}(x)\right]
-\frac{1}{2}U_1(x)[\psi_{+2}(x+\hat{1}),\psi_{-2}(x+\hat{1})]
U_1(x)^{-1} \nn \\
&\qquad\qquad
-\psi_{+1}(x)U_1(x)\psi_{-2}(x+\hat{1})U_1(x)^{-1}
+\psi_{-1}(x)U_1(x)\psi_{+2}(x+\hat{1})U_1(x)^{-1}
\Bigr\} U_{12}(x) \nn \\
&\qquad
+U_{12}(x)\Bigl\{
-\frac{1}{2}\left[\psi_{+2}(x),\psi_{-2}(x)\right]
-\frac{1}{2}U_2(x)[\psi_{+1}(x+\hat{2}),\psi_{-1}(x+\hat{2})]
U_2(x)^{-1} \nn \\
&\qquad\qquad\qquad\quad
-U_2(x)\psi_{+1}(x+\hat{2})U_2(x)^{-1}\psi_{-2}(x)
+U_2(x)\psi_{-1}(x+\hat{2})U_2(x)^{-1}\psi_{+2}(x)
\Bigr\} \nn \\
&\qquad
-\Bigl\{
-\frac{1}{2}\left[\psi_{+2}(x),\psi_{-2}(x)\right]
-\frac{1}{2}U_2(x)[\psi_{+1}(x+\hat{2}),\psi_{-1}(x+\hat{2})]
U_2(x)^{-1} \nn \\
&\qquad\qquad
-\psi_{+2}(x)U_2(x)\psi_{-1}(x+\hat{2})U_2(x)^{-1}
+\psi_{-2}(x)U_2(x)\psi_{+1}(x+\hat{2})U_2(x)^{-1}
\Bigr\}U_{21}(x) \nn \\
&\qquad
-U_{21}(x)\Bigl\{
-\frac{1}{2}\left[\psi_{+1}(x),\psi_{-1}(x)\right]
-\frac{1}{2}U_1(x)[\psi_{+2}(x+\hat{1}),\psi_{-2}(x+\hat{1})]
U_1(x)^{-1} \nn \\
&\qquad\qquad\qquad\quad
-U_1(x)\psi_{+2}(x+\hat{1})U_1(x)^{-1}\psi_{-1}(x)
+U_1(x)\psi_{-2}(x+\hat{1})U_1(x)^{-1}\psi_{+1}(x)
\Bigr\} \nn \\
&\qquad
-\bigl(\psi_{-1}(x)+U_1(x)\psi_{-2}(x+\hat{1})U_1(x)^{-1}\bigr)U_{12}(x)
 \nn \\\
&\hspace{6cm}
\times\bigl(\psi_{+2}(x)+U_2(x)\psi_{+1}(x+\hat{2})U_2(x)^{-1}\bigr)
 \nn \\
&\qquad
+\bigl(\psi_{+1}(x)+U_1(x)\psi_{+2}(x+\hat{1})U_1(x)^{-1}\bigr)U_{12}(x)
 \nn \\\
&\hspace{6cm}
\times\bigl(\psi_{-2}(x)+U_2(x)\psi_{-1}(x+\hat{2})U_2(x)^{-1}\bigr)
 \nn \\
&\qquad
+\bigl(\psi_{-2}(x)+U_2(x)\psi_{-1}(x+\hat{2})U_2(x)^{-1}\bigr)U_{21}(x)
 \nn \\\
&\hspace{6cm}
\times\bigl(\psi_{+1}(x)+U_1(x)\psi_{+2}(x+\hat{1})U_1(x)^{-1}\bigr)
 \nn \\
&\qquad
-\bigl(\psi_{+2}(x)+U_2(x)\psi_{+1}(x+\hat{2})U_2(x)^{-1}\bigr)U_{21}(x)
 \nn \\\
&\hspace{6cm}
\times\bigl(\psi_{-1}(x)+U_1(x)\psi_{-2}(x+\hat{1})U_1(x)^{-1}\bigr), 
\\
&\Tr\Bigl[
Q_+Q_- \left(U_{12}(x)+U_{21}(x)\right)
\Bigr]\Bigr|_{\rm fermion} 
\nn \\
&\qquad 
=\Tr\Biggl[
U_{12}(x)\biggl\{
-\left\{\psi_{+1}(x),\psi_{-2}(x)\right\}
+\left\{\psi_{-1}(x),\psi_{+2}(x)\right\} \nn \\
&\hspace{2.8cm}
+\left(\cD_2\psi_{+1}(x)\right)\left(\cD_1\psi_{-2}(x)\right) 
-\left(\cD_2\psi_{-1}(x)\right)\left(\cD_1\psi_{+2}(x)\right) \nn \\
&\hspace{2.8cm}
+\frac{1}{2}\left(\cD_2\psi_{+1}(x)\right) \psi_{-1}(x)
+\frac{1}{2}U_2(x)\psi_{-1}(x+\hat{2})U_2(x)^{-1} \left(\cD_2\psi_{+1}(x)\right) \nn \\
&\hspace{2.8cm}
-\frac{1}{2}\left(\cD_2\psi_{-1}(x)\right) \psi_{+1}(x)
-\frac{1}{2}U_2(x)\psi_{+1}(x+\hat{2})U_2(x)^{-1}  \left(\cD_2\psi_{-1}(x)\right) \nn \\
&\hspace{2.8cm}
+\frac{1}{2}\psi_{+2}(x) \left(\cD_1\psi_{-2}(x)\right)
+\frac{1}{2}\left(\cD_1\psi_{-2}(x)\right)
 U_1(x)\psi_{+2}(x+\hat{1})U_1(x)^{-1} \nn \\
&\hspace{2.8cm}
-\frac{1}{2}\psi_{-2}(x) \left(\cD_1\psi_{+2}(x)\right) 
-\frac{1}{2}\left(\cD_1\psi_{+2}(x)\right)
 U_1(x)\psi_{-2}(x+\hat{1})U_1(x)^{-1} 
\biggr\} \nn \\
&\qquad\qquad
+U_{21}(x)\biggl\{
\left\{\psi_{+1}(x),\psi_{-2}(x)\right\}
-\left\{\psi_{-1}(x),\psi_{+2}(x)\right\} \nn \\
&\hspace{2.8cm}
+\left(\cD_1\psi_{+2}(x)\right)\left(\cD_2\psi_{-1}(x)\right)
-\left(\cD_1\psi_{-2}(x)\right)\left(\cD_2\psi_{+1}(x)\right) \nn \\
&\hspace{2.8cm}
+\frac{1}{2}\psi_{+1}(x) \left(\cD_2\psi_{-1}(x)\right)
+\frac{1}{2}\left(\cD_2\psi_{-1}(x)\right) U_2(x)\psi_{+1}(x+\hat{2})U_2(x)^{-1} \nn \\
&\hspace{2.8cm}
-\frac{1}{2} \psi_{-1}(x) \left(\cD_2\psi_{+1}(x)\right)
-\frac{1}{2} \left(\cD_2\psi_{+1}(x)\right) U_2(x)\psi_{-1}(x+\hat{2})U_2(x)^{-1} \nn \\
&\hspace{2.8cm}
+\frac{1}{2}\left(\cD_1\psi_{+2}(x)\right) \psi_{-2}(x)
+\frac{1}{2} U_1(x)\psi_{-2}(x+\hat{1})U_1(x)^{-1} \left(\cD_1\psi_{+2}(x)\right) \nn \\
&\hspace{2.8cm}
-\frac{1}{2}\left(\cD_1\psi_{-2}(x)\right) \psi_{+2}(x)
-\frac{1}{2} U_1(x)\psi_{+2}(x+\hat{1})U_1(x)^{-1} \left(\cD_1\psi_{-2}(x)\right) 
\biggr\}
\Biggr], \\
&-iQ_+ \left(U_{12}(x)-U_{21}(x)\right)
\Tr\Bigl[Q_-\left(U_{12}(x)+U_{21}(x)\right)\Bigr]  \nn \\
&\quad +iQ_- \left(U_{12}(x)-U_{21}(x)\right)
\Tr\Bigl[Q_+\left(U_{12}(x)+U_{21}(x)\right)\Bigr]
\nn \\
&\qquad =\biggl\{
\bigl(\psi_{+1}(x)+U_1(x)\psi_{+2}(x+\hat{1})U_1(x)^{-1}\bigr)U_{12}(x)
 \nn \\
&\hspace{1.7cm}
-U_{12}(x)\bigl(\psi_{+2}(x)+U_2(x)\psi_{+1}(x+\hat{2})U_2(x)^{-1}\bigr)
 \nn \\ 
&\hspace{1.7cm}
-\bigl(\psi_{+2}(x)+U_2(x)\psi_{+1}(x+\hat{2})U_2(x)^{-1}\bigr)U_{21}(x)
 \nn \\
&\hspace{1.7cm}
+U_{21}(x)\bigl(\psi_{+1}(x)+U_1(x)\psi_{+2}(x+\hat{1})U_1(x)^{-1}\bigr) 
\biggr\} \nn \\
&\hspace{1.5cm}
\times\Tr\Bigl[
i\left(U_{12}(x)-U_{21}(x)\right)\Bigl(
-\cD_2\psi_{-1}(x) + \cD_1\psi_{-2}(x)
\Bigr)
\Bigr] \nn \\
&\qquad 
-\biggl\{
\bigl(\psi_{-1}(x)+U_1(x)\psi_{-2}(x+\hat{1})U_1(x)^{-1}\bigr)U_{12}(x)
 \nn \\
&\hspace{1.7cm}
-U_{12}(x)\bigl(\psi_{-2}(x)+U_2(x)\psi_{-1}(x+\hat{2})U_2(x)^{-1}\bigr)
 \nn \\ 
&\hspace{1.7cm}
-\bigl(\psi_{-2}(x)+U_2(x)\psi_{-1}(x+\hat{2})U_2(x)^{-1}\bigr)U_{21}(x)
 \nn \\
&\hspace{1.7cm}
+U_{21}(x)\bigl(\psi_{-1}(x)+U_1(x)\psi_{-2}(x+\hat{1})U_1(x)^{-1}\bigr) 
\biggr\} \nn \\
&\hspace{1.5cm}
\times\Tr\Bigl[
i\left(U_{12}(x)-U_{21}(x)\right)\Bigl(
-\cD_2\psi_{+1}(x) + \cD_1\psi_{+2}(x)
\Bigr)
\Bigr], 
\\
&\Tr\Bigl[Q_+\left(U_{12}(x)+U_{21}(x)\right)\Bigr]
\Tr\Bigl[Q_-\left(U_{12}(x)+U_{21}(x)\right)\Bigr] \nn \\
&\qquad =
\Tr\Bigl[i\left(U_{12}(x)-U_{21}(x)\right)
\Bigl(-\cD_2\psi_{+1}(x)+\cD_1\psi_{+2}(x)\Bigr)\Bigr] \nn \\
&\qquad
\times\Tr\Bigl[i\left(U_{12}(x)-U_{21}(x)\right)
\Bigl(-\cD_2\psi_{-1}(x)+\cD_1\psi_{-2}(x)\Bigr)\Bigr]. 
\end{align}


\providecommand{\href}[2]{#2}\begingroup\raggedright\endgroup

\end{document}